\documentclass[prb,aps,amssymb,10pt,floatfix,showpacs,notitlepage, twocolumn, superscriptaddress]{revtex4-2}
\usepackage{bbm}
\usepackage{amsmath}
\usepackage{amssymb}
\usepackage{tensor}
\usepackage{color,graphicx}
\graphicspath{{figures/}{suppmat/figures/}}
\usepackage{hyperref}
\usepackage{interval}
\usepackage{tikz}
\usepackage{afterpage}
\usepackage{multirow}
\usepackage[titles]{tocloft}

\newcommand{\ket}[1]{|#1\rangle}
\newcommand*{\bra}[1]{\langle#1|}

\newcommand*{\myexp}[1]{\left\langle#1\right\rangle}
\newcommand{\ie}{\textit{i.e.~}}
\newcommand{\wrt}{\textit{w.r.t.~}}
\DeclareMathOperator{\tr}{\mathrm{Tr}}
\newcommand{\cf}{{\it cf.~}}

\newcommand{\cfourz}{C_4^z}

\newcommand{\reftosuppmat}[1]{(see App.~\ref{#1})}

\DeclareMathOperator{\swap}{\mathrm{SWAP}}
\DeclareMathOperator{\swapAmp}{\swap_{\text{amp}}}
\DeclareMathOperator{\swapPhase}{\swap_{\text{phase}}}
\DeclareMathOperator{\var}{\mathrm{Var}}

\begin{document}
	
	\title{Fractional Chiral Hinge Insulator}
	
	\author{Anna Hackenbroich}
	
	\affiliation{Max-Planck-Institute of Quantum Optics, Hans-Kopfermann-Stra{\ss}e 1, 85748 Garching, Germany}
	
	\affiliation{Munich Center for Quantum Science and Technology, Schellingstra{\ss}e 4, 80799 M{\"u}nchen, Germany}
	
	\author{Ana Hudomal}
	
	\affiliation{Institute of Physics Belgrade, University of Belgrade, 11080 Belgrade, Serbia}	
	\affiliation{School of Physics and Astronomy, University of Leeds, Leeds, LS2 9JT, United Kingdom}
	
	\author{Norbert Schuch}
	
	\affiliation{Max-Planck-Institute of Quantum Optics, Hans-Kopfermann-Stra{\ss}e 1, 85748 Garching, Germany}
	
	\affiliation{Munich Center for Quantum Science and Technology, Schellingstra{\ss}e 4, 80799 M{\"u}nchen, Germany}
	
	\affiliation{University of Vienna, Department of Physics, Boltzmanngasse 5, 1090 Wien, Austria}
	\affiliation{University of Vienna, Department of Mathematics, Oskar-Morgenstern-Platz 1, 1090 Wien, Austria}
	
	\author{B. Andrei Bernevig}
	
	\affiliation{Joseph Henry Laboratories and Department of Physics, Princeton University, Princeton, New Jersey 08544, USA}
	
	\author{Nicolas Regnault}
	\affiliation{Joseph Henry Laboratories and Department of Physics, Princeton University, Princeton, New Jersey 08544, USA}
	
	\affiliation{Laboratoire de Physique de l'Ecole normale sup\'{e}rieure, ENS, Universit\'{e} PSL, CNRS, Sorbonne Universit\'{e}, Universit\'{e} Paris-Diderot, Sorbonne Paris Cit\'{e}, Paris, France}

\begin{abstract}
	We propose and study a wave function describing an interacting three-dimensional fractional chiral hinge insulator (FCHI) constructed by Gutzwiller projection of two non-interacting second order topological insulators with chiral hinge modes at half filling. We use large-scale variational Monte Carlo computations to characterize the model states via the entanglement entropy and charge-spin-fluctuations. We show that the FCHI possesses fractional chiral hinge modes characterized by a central charge $c=1$ and Luttinger parameter $K=1/2$, like the edge modes of a Laughlin $1/2$ state. By changing the boundary conditions for the underlying fermions, we investigate the topological degeneracy of the FCHI. Within the range of the numerically accessible system sizes, we observe a non-trivial topological degeneracy. A more numerically pristine characterization of the bulk topology is provided by the topological entanglement entropy (TEE) correction to the area law. While our computations indicate a vanishing bulk TEE, we show that the gapped surfaces host a two-dimensional topological order with a TEE per surface compatible with half that of a Laughlin $1/2$ state, a value that cannot be obtained from topological quantum field theory. 
\end{abstract}

	\maketitle

\addtocontents{toc}{\protect\setcounter{tocdepth}{0}}
\addtocontents{lot}{\protect\setcounter{lotdepth}{-1}}
\addtocontents{lof}{\protect\setcounter{lofdepth}{-1}}
		
	\section{Introduction}
	
	Strong interactions in condensed matter systems can lead to fascinating emergent phenomena. In two-dimensional (2D) systems, strong interactions may lead to the emergence of topological order (TO), such as experimentally observed in the fractional quantum Hall effect. Features of TO in 2D include a non-trivial ground state degeneracy on certain surfaces and the appearance of itinerant excitations with fractional quantum numbers and braiding statistics. It has long been an active field of study to extend this rich physics to three-dimensional (3D) strongly interacting systems, where the emergent physics can be even more diverse, including systems with fractonic excitations~\cite{doi:10.1146/annurev-conmatphys-031218-013604,doi:10.1142/S0217751X20300033}. Whereas many microscopic models based on interacting spin systems have been proposed to exhibit TO in 3D, such as the 3D toric code~\cite{PhysRevB.78.155120} and 3D Kitaev models~\cite{PhysRevB.79.075124, Mondragon_Shem_2014, PhysRevB.79.024426, PhysRevB.98.125136}, there is a scarcity of electronic or realistic examples that could be experimentally relevant.
	
Among the 3D electronic topological insulators (TIs), an entirely new class has recently been discovered: certain TIs protected by crystalline symmetries, now dubbed higher order TIs~\cite{benalcazar2017quantizedScience, benalcazar2017quantized, schindler2018higher, PhysRevLett.119.246402, PhysRevLett.119.246401, SiddharthCorner, PhysRevB.97.205136, PhysRevB.97.205135, PhysRevX.9.011012, PhysRevB.98.235102, PhysRevB.101.085137, PhysRevB.100.020509, PhysRevB.99.235132, 2020arXiv200514500Z, 2020PhRvR...2c3192Y, 2020arXiv200313706K}, possess a much richer bulk-boundary correspondence than conventional, or first order, TIs. For example, there exists a 3D chiral hinge insulator (CHI), whose gapped surfaces are connected by gapless chiral hinge modes~\cite{schindler2018higher}. Higher order TIs in two and three dimensions have been experimentally observed in either materials~\cite{SchindlerBismuth}, mechanical~\cite{serra2018observation}, acoustic~\cite{HoaranAcousticHOTI, XiangAcoustic2Dchiral}, photonic~\cite{mittal2018photonic, hassan2018cornerphotonic, xie2018visualizationphotonic, yang2019gapped} or electrical~\cite{peterson2018quantized,imhof2018topolectrical,PhysRevB.99.020304} systems. 

In this letter, we provide a first stepping stone in the realization of a full-fledged electronic 3D fractional TI by building a 3D fractional chiral hinge insulator (FCHI) model wave function. Indeed, the hinge modes of the non-interacting CHI are of the same nature as the edge modes of a Chern insulator, two copies of which at fractional filling and with strong interactions form a fractional Chern insulator (FCI) hosting fractional quantum Hall physics~\cite{sheng2011fractional, PhysRevLett.106.236804, PhysRevX.1.021014}. Therefore, we may speculate that under similar conditions the FCHI will also display non-trivial topology with fractionalized excitations at least at the hinges or surfaces. The FCHI could also represent another lane between higher order TI and fractonic systems~\cite{2019arXiv190905868Y}.

Numerical computations and especially exact diagonalizations for interacting electronic systems in 3D are notoriously difficult due to the spatial dimensionality. To partially circumvent this challenge, we will rely on a model wave function, a fruitful approach for TO, to capture the FCHI. This approach has been extensively applied in the realm of the fractional quantum Hall effect~\cite{PhysRevLett.50.1395, MOORE1991362} and FCIs~\cite{PhysRevB.87.161113}. In order to define the FCHI wave function, we will make use of Gutzwiller projection, a systematic method to construct interacting model wave functions starting from copies of non-interacting ground states. Large-scale variational Monte Carlo (MC) simulations then allow us to analyze this  wave function for bigger system sizes than possible with other methods.

To probe the topological content of the wave function, we will study the entanglement entropy (EE), which can be evaluated in MC simulations~\cite{PhysRevLett.114.206402,PhysRevB.84.075128, PhysRevB.87.161113}, and follows an area law with characteristic subleading corrections~\cite{PhysRevLett.71.666}. In two dimensions there are logarithmic corrections for gapless edge modes~\cite{Calabrese_2009, crepel2019model, crepel2019microscopic} which along with the constant topological entanglement entropy (TEE) correction to the bulk area law~\cite{PhysRevLett.96.110404, PhysRevLett.96.110405} provide information on the system's topology. In three dimensions, corrections to the bulk area law include the TEE and possible size-dependent corrections for fractonic systems and layered constructions~\cite{PhysRevB.84.195120, PhysRevB.97.125101, PhysRevB.97.125102}. In particular, we study the hinge modes in an open system and show that they are fractionalized excitations characterized by a central charge $c=1$ and Luttinger parameter $K=1/2$, like the FCI edge modes. We then study the linear independence of different interacting wave functions obtained by changing the boundary conditions for the underlying fermions, thus finding a non-trivial topological degeneracy for the numerically accessible system sizes. Finally, we study the TEE of the bulk system, and that of the gapped surfaces. Whereas our computations indicate a vanishing bulk TEE, we show that the gapped surfaces host a non-trivial two-dimensional topological phase with a TEE per surface compatible with half that of a Laughlin $1/2$ state.
	
	\section{Model wave function}
	
\begin{figure}[t]	
	\includegraphics[width = \linewidth] {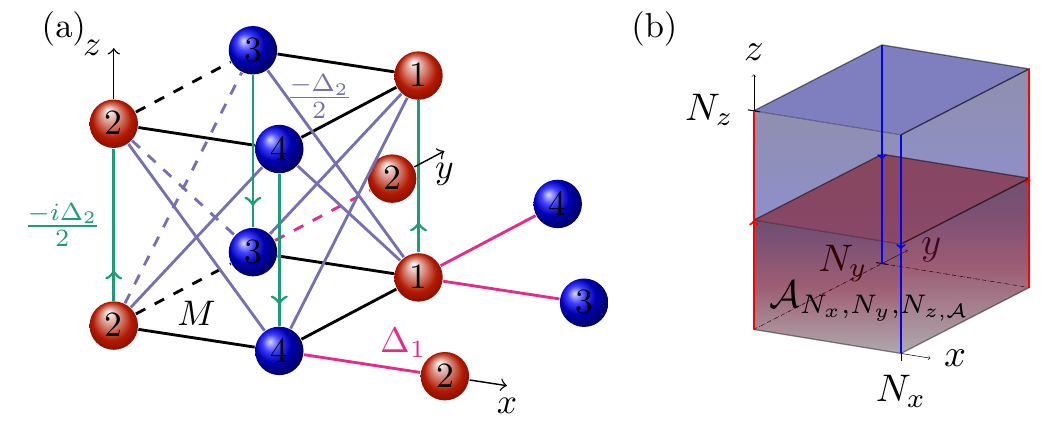}
	
	\caption[Sketch CHI model]{(a) Local real-space model for a 3D second order TI with chiral hinge states. The Hamiltonian is defined on a cubic lattice with a unit cell of four sites lying in the $xy$-plane. In this plane, sites in the same unit cell are connected by a nearest-neighbour hopping $M$ marked by black lines ($-M$ for dashed black lines). In the $xy$-plane, sites in adjacent unit cells are connected by a nearest-neighbour hopping $\Delta_1$ marked by violet lines ($-\Delta_1$ for dashed violet lines). In the $z$ direction, adjacent unit cells are connected by a real next-nearest neighbour hopping $- \Delta_2/ 2$ marked by light blue lines ($\Delta_2/ 2$ for dashed light blue lines). In addition, there is a purely imaginary nearest neighbour hopping between adjacent unit cells in the $z$ direction with value $-i\Delta_2 / 2$ in the direction of the green arrows. We study the model for parameter values $M = \Delta_1 = \Delta_2 = 1$, where the correlation length is close to its minimal value~\reftosuppmat{sec:CHI}. (b) 3D system with OBC and $N_x, N_y$ unit cells in the $x, y$ directions, and periodic boundaries and $N_z$ sites in the $z$ direction. The subsystem $\mathcal{A}_{N_{x}, N_{y}, N_{z, \mathcal{A}}}$ consists of $N_{x}, N_{y}$ unit cells in the $x, y$ directions and $N_{z, \mathcal{A}}$ unit cells in the $z$ direction.\label{fig:SketchCHI}}
\end{figure}

	We consider an interacting model wave function obtained by Gutzwiller projection of the ground state of a non-interacting 3D second-order TI with chiral hinge modes. The CHI model is described by a local Hamiltonian for spinless fermions with four sites per unit cell~\cite{schindler2018higher} (see Fig.~\ref{fig:SketchCHI} (a) for a sketch of the model). The ground state $\ket{\psi}$ of the CHI model lies at filling $\nu = 1/2$ of the lattice. With open boundary conditions (OBC) in the $x$ and $y$ directions, each of the four hinges of the CHI parallel to the $z$-axis supports a single chiral mode localized at the hinge. Each hinge mode corresponds to a free bosonic mode with central charge $c=1$ and Luttinger parameter $K=1$ akin to the edge modes of a Chern insulator~\reftosuppmat{sec:CHI}. Since the CHI model is non-interacting, it does not have TO or a non-trivial ground state degeneracy with periodic boundary conditions (PBC).
	
	In order to define the interacting model wave function $\ket{\Psi}$, we take two copies $\ket{\psi_{s}}$ of the ground state of the CHI model at half filling, to which we assign different values $ s \in \{ \uparrow, \downarrow\}$ of a spin-like degree of freedom. The interacting wave function is obtained as the Gutzwiller projection
	\begin{equation}
	\ket{\Psi} = P_G \left[ \ket{\psi_{\uparrow}} \otimes \ket{\psi_{\downarrow}} \right]
	\end{equation} 
	of the product of the two non-interacting wave functions. With $\hat{n}_{s,i}$ denoting the particle number operator for fermions of spin $s$ on the lattice site $i$, the Gutzwiller projection operator is expressed as
	\begin{equation}
	P_G = \prod_{i} (1 - \hat{n}_{\uparrow, i} \hat{n}_{\downarrow, i}).
	\end{equation}
	It forbids simultaneous occupancy of any lattice site $i$ by both a particle with spin $\uparrow$ and spin $\downarrow$. Therefore, it simulates the effect of a very large on-site Hubbard interaction. Since each copy of the ground state of the CHI model has a filling $\nu_{\psi_{\uparrow}} = \nu_{\psi_{\downarrow}} = 1/2$, the Gutzwiller projection enforces that the interacting wave function lies at filling $\nu_{\Psi} = 1/2$ with exactly one particle per lattice site (each lattice site having a spin degree of freedom which can take two values). Hence, charge fluctuations are completely frozen and the only relevant degree of freedom in the interacting wave function is the spin $s$.

	\section{Characterization of hinge modes}
	
	With OBC in the $x$ and $y$ directions, the interacting model wave function $\ket{\Psi}$ is expected to posses one gapless chiral mode at each of the four hinges parallel to the $z$-axis, inherited from the hinge modes of the non-interacting CHI. Like the edge modes of chiral topologically ordered phases in two dimensions, we expect the hinge modes of $\ket{\Psi}$ to be described by a chiral conformal field theory (CFT). Moreover, since $\ket{\Psi}$ is interacting, we expect its hinge CFT to be possibly different than the trivial free-boson CFT describing the hinge modes of the non-interacting CHI.
	
	In order to characterize the chiral hinge modes, we adapt the methods that have previously been employed for 2D chiral phases~\cite{crepel2019model, crepel2019microscopic,estienne2019entanglement} to the 3D setting: We study the second Renyi entropy $S^{(2)}$ and spin fluctuations of $\ket{\Psi}$, in focusing on the critical contributions stemming from the \emph{physical} hinges. We evaluate these observables for the interacting wave function $\ket{\Psi}$ in large-scale MC simulations using the $\swap$-operator technique~\cite{PhysRevLett.104.157201} with sign-problem refinement~\cite{PhysRevLett.107.067202}~\reftosuppmat{sec:AppendixMC}. 
	
	We consider the geometry sketched in Fig.~\ref{fig:SketchCHI}(b): A total system with $N_x \times N_y \times N_z$ unit cells, OBC in the $xy$-plane, and PBC in the $z$ direction to ensure that the only gapless excitations are the four hinge states. We consider a series of subsystems $\mathcal{A}_{N_x, N_y, N_{z, \mathcal{A}}}$ with $N_x, N_y$ unit cells in the $x, y$ directions and $N_{z, \mathcal{A}} \in \{1, \dotsc, N_z-1\}$ unit cells in the $z$ direction, marked in red in Fig.~\ref{fig:SketchCHI}(b). The $\mathcal{A}_{N_x, N_y, N_{z, \mathcal{A}}}$ bisect each of four physical hinge modes into a part of length $ N_{z, \mathcal{A}}$ contained in  $\mathcal{A}_{N_x, N_y, N_{z, \mathcal{A}}}$, and the remaining part outside of the subsystem. Hence, we expect that the EE and spin fluctuations \wrt $\mathcal{A}_{N_x, N_y, N_{z, \mathcal{A}}}$ will contain signatures from the hinges.

	Specifically, if the hinge modes are described by a chiral CFT with central charge $c$, the second Renyi entropy $S^{(2)}$ of $\ket{\Psi}$ \wrt the $\mathcal{A}_{N_x, N_y, N_{z, \mathcal{A}}}$ for different $N_{z, \mathcal{A}}$ at fixed $N_x$ and $N_y$ is expected to scale as 
	\begin{equation}\label{EEHinge}
	S^{(2)}_{\mathcal{A}_{N_x, N_y, N_{z, \mathcal{A}}}} (N_{z, \mathcal{A}}) = \alpha + 4 \times S^{(2)}_{\text{crit}} (N_{z, \mathcal{A}}; N_z). 
	\end{equation}
	Here, $\alpha$ is a constant independent of $N_{z, \mathcal{A}}$. It includes the area law contributions from the virtual surfaces at $z = 0, N_{z, \mathcal{A}}$ which scale proportional to $N_xN_y$, and are therefore independent of $N_{z, \mathcal{A}}$ in the thermodynamic limit, and any potential corner contributions. In Eq.~\eqref{EEHinge}, 	
	\begin{equation}\label{EE1DCritical}
	S^{(2)}_{\text{crit}} (N_{z, \mathcal{A}}; N_z) = \frac{c}{8} \ln \left[ \frac{N_z}{\pi} \sin \left( \frac{\pi N_{z, \mathcal{A}}}{N_z}\right)\right]
	\end{equation}
	is the second Renyi entropy of a periodic one-dimensional chiral critical mode with central charge $c$ and total system size $N_z$ restricted to a single interval of length $N_{z, \mathcal{A}}$~\cite{Calabrese_2009}. The factor of $4$ in Eq.~\eqref{EEHinge} takes into account the four hinge modes, which contribute equally to the EE.
	
\begin{figure}[t]

	\includegraphics[width = \linewidth] {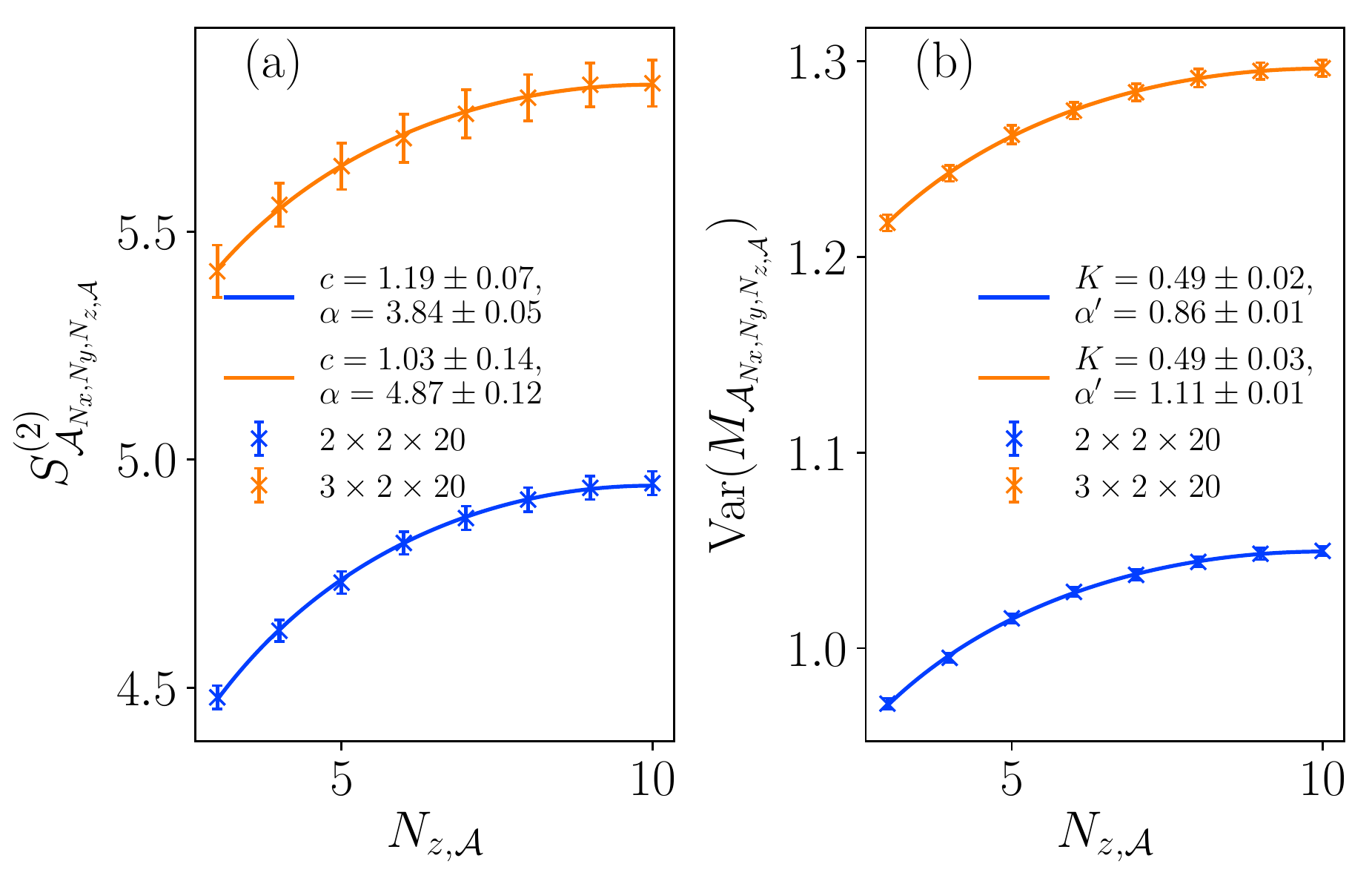}
	\caption[Hinge mode characterization FCHI]{Second Renyi entropy and spin fluctuations of the interacting model wave function $\ket{\Psi}$ for a series of subsystems $\mathcal{A}_{N_x, N_y, N_{z, \mathcal{A}}}$ (for a sketch see Fig.~\ref{fig:SketchCRealSpaceCI}(b)). We plot MC data obtained for two different systems sizes $2 \times 2 \times 20$ (in blue) and $3 \times 2 \times 20$ (in orange). (a) Scaling of the second Renyi entropy, fit to the prediction of Eq.~\eqref{EEHinge}. (b) Scaling of the spin fluctuations, fit to the prediction of Eq.~\eqref{FitFunctionParticleNumberVariance3D}.\label{fig:EEChiralHingeTi}}
	\end{figure}

	The scaling of the second Renyi entropy of $\ket{\Psi}$ as computed from MC is shown in Fig.~\ref{fig:EEChiralHingeTi}(a) for two different system sizes $2 \times 2 \times 20$ and $3 \times 2 \times 20$. For computational reasons, we choose $N_x$ and $N_y$ much smaller than $N_z$~\reftosuppmat{sec:CHI}. Due to the short correlation length of the CHI, equal to one lattice spacing~\reftosuppmat{sec:CHI}, we may expect that the characteristic parameters approach their thermodynamic limit even for small $N_x, N_y$. The logarithmic scaling from the hinge states is clearly visible, and numerical values for $c$ and $\alpha$ can be extracted by fitting the data to Eq.~\eqref{EEHinge}. The numerical value for the central charge is $c = 1.19 \pm 0.07$ for $2 \times 2 \times 20$ and $c = 1.03 \pm 0.14$ for $3 \times 2 \times 20$. This provides strong evidence that the hinge modes of the interacting model wave function $\ket{\Psi}$ are described by a chiral free-boson CFT with central charge $c = 1$.

	Free-boson CFTs with $c = 1$ are characterised by their Luttinger parameter $K$. For such Luttinger liquids, the variance of the U(1) current integrated over a subsystem scales proportionally to the EE, where the proportionality constant allows the extraction of $K$~\cite{estienne2019entanglement}. Since charge fluctuations are completely frozen in the wave function $\ket{\Psi}$, the relevant U(1) symmetry stems from the spin degree of freedom, and we need to consider the fluctuations of the number $M_{\mathcal{A}}$ of particles with spin $\uparrow$ in a subsystem $\mathcal{A}$. Concretely, we consider the variance 	
	\begin{equation}
	\var(M_{\mathcal{A}_{N_x, N_y, N_{z, \mathcal{A}}}}) \equiv \langle M_{\mathcal{A}_{N_x, N_y, N_{z, \mathcal{A}}}}^2 \rangle - \langle M_{\mathcal{A}_{N_x, N_y, N_{z, \mathcal{A}}}} \rangle ^2.
	\end{equation}
	which is expected to scale as~\cite{estienne2019entanglement}  	
	\begin{equation}\label{FitFunctionParticleNumberVariance3D}
	\var(M_{\mathcal{A}_{N_x, N_y, N_{z, \mathcal{A}}}}) = 2 \times \frac{K} {\pi ^ 2}  \ln \left[ \frac{N_z}{\pi} \sin \left(\frac{\pi  N_{z, \mathcal{A}}} {N_z} \right)\right] + \alpha'
	\end{equation}	
	with the Luttinger parameter $K$ and a constant $\alpha'$ independent of $N_{z, \mathcal{A}}$. 
	
	The scaling of the spin fluctuations in the wave function $\ket{\Psi}$ as computed from MC is shown in Fig.~\ref{fig:EEChiralHingeTi}(b) for two different system sizes $2 \times 2 \times 20$ and $3 \times 2 \times 20$. Remarkably, even for these small sizes, the numerical value for $K$ extracted by fitting the data to Eq.~\eqref{FitFunctionParticleNumberVariance3D} is $K = 0.49\pm 0.02$ for $2 \times 2 \times 20$ and $K = 0.49\pm 0.03$ for $3 \times 2 \times 20$. This provides strong evidence that the Luttinger parameter for the chiral hinge modes of the interacting higher order TI is $K = 1/2$, similarly to the edge modes of a FCI.

	\section{Topological degeneracy and topological entanglement entropy}
	
	In two dimensions, fractionalized excitations such as those of the edge modes of an FCI are an indication for bulk TO. Above, we showed that the FCHI has fractional hinge modes. It is therefore natural to investigate if it also possesses non-trivial topology in the bulk.
	
	2D topologically ordered systems are characterized by a non-zero TEE and a non-trivial topological degeneracy on surfaces with a genus greater than zero. In three dimensions, TEE and topological degeneracy remain important bulk signatures of non-trivial topology and can display various forms. For example, 3D systems with intrinsic TO have a ground state degeneracy which depends only on the topology of the space, such as the 3D Kitaev model, which has a topological degeneracy of 8 on the 3-torus~\cite{PhysRevB.90.104424}. On the other hand, fractonic systems possess a ground state degeneracy which might grow exponentially with the system size~\cite{doi:10.1142/S0217751X20300033}. They can also exhibit non-trivial corrections to the area law, which are also size-dependent~\cite{PhysRevB.84.195120, PhysRevB.97.125101, PhysRevB.97.125102}.
	
	\paragraph{Topological degeneracy}
	
	In order to study the topological degeneracy of the FCHI we closely follow a well-known approach established for 2D projected wave functions such as the FCI. On the 2D torus, one defines four interacting wave functions by choosing PBC or anti-periodic boundary conditions (APBC) for the underlying fermions in each direction of the torus. For the FCI, these four states yield two linearly independent wave functions as expected in the phase of the Laughlin wave function with filling $\nu = 1 / 2$~\reftosuppmat{sec:FCI_TopDeg}.  
	
	For the FCHI, we consider 8 independent ansatz states on the 3D torus obtained by Gutzwiller projection of the non-interacting CHI wave function with PBC or APBC in each direction. The ground state degeneracy is then given by the rank of an 8-dimensional overlap matrix $\mathcal{O}$ containing the normalized overlaps of these ansatz states~\reftosuppmat{sec:AppendixFCHI}. Note that the topological degeneracy could in principle be larger than 8, in particular for fractonic systems. In such a case, the rank of the overlap matrix considered here would still be at most 8 and our approach would fail to measure the full ground state degeneracy. 
	
\begin{figure}[t]	
	\includegraphics[width = \linewidth] {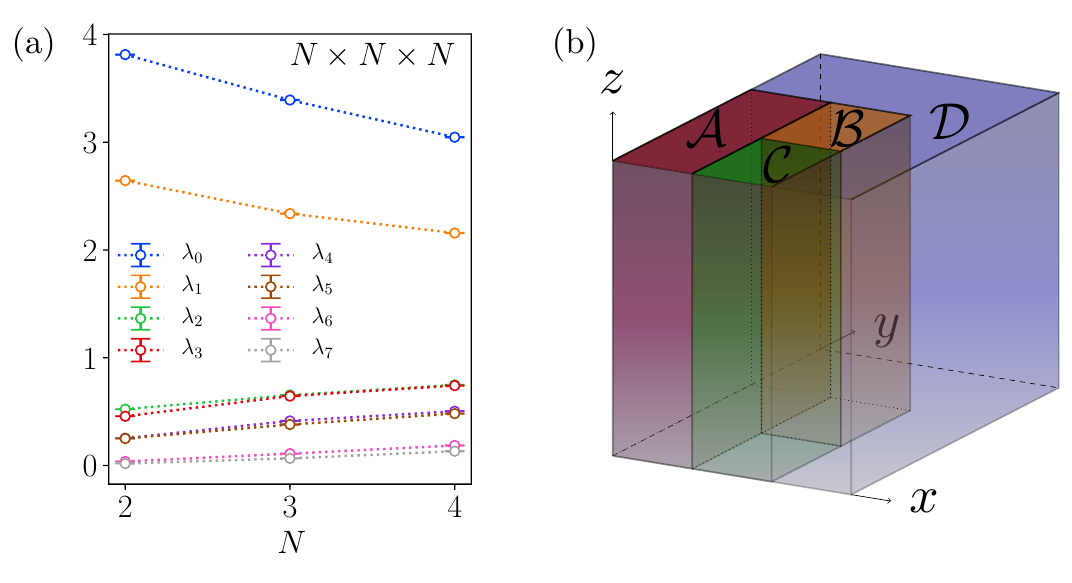}
	\caption[3D Kitaev-Preskill cut]{(a) Scaling of the eigenvalues $\lambda_i$ with $i = 0, \dotsc, 7$ of the overlap matrix $\mathcal{O}$ of the FCHI on the isotropic 3-torus with $N \times N \times N$ unit cells. (b) Subsystems $\mathcal{A}$, $\mathcal{B}$, $\mathcal{C}$ and $\mathcal{D}$ for the extraction of the bulk TEE using a Kitaev-Preskill cut. Note that the subsystems are translation invariant in the $z$-direction.\label{fig:SketchKitaevPreskill3D}}
	\end{figure}

	We have studied the topological degeneracy of the FCHI on isotropic 3-tori with $N \times N\times N$ unit cells up to $N = 4$ using variational MC simulations~\reftosuppmat{sec:AppendixFCHI}. The results are shown in Fig.~\ref{fig:SketchKitaevPreskill3D}(a). For these system sizes, we observe a separation of the eigenvalues of the overlap matrix $\mathcal{O}$ into a group of two larger eigenvalues and a group of 6 smaller eigenvalues. However, there is no clear trend indicating that the former would converge to a finite value and the latter to zero in the thermodynamic limit.
	
	On the other hand, for very anisotropic 3-tori with $N_z$ much larger than $N_x=N_y$ we observe a clear separation of the eigenvalues of $\mathcal{O}$ into a group of four large and a group of four small eigenvalues, which approach the values $2$ and $0$, respectively, \emph{exponentially} fast as a function of growing $N_z$ for $N_x$ and $N_y$ constant. In the limit $N_z \rightarrow \infty$, we thus find four linearly independent wave functions associated with the four different boundary conditions in the horizontal directions, whereas the system becomes insensitive to the boundary conditions in the $z$ direction. This is tightly related to the behavior of the underlying non-interacting wave function, for which the normalised overlap between two many-body wave functions corresponding to different boundary conditions in the $z$ direction also approaches unity as $N_z \rightarrow \infty$ for $N_x$ and $N_y$ constant. However, in the non-interacting system this approach is only algebraic as a function of $N_z$, whereas it is exponential in the interacting system~\reftosuppmat{sec:AppendixFCHI}. 
	
	\paragraph{Topological entanglement entropy}
	
	In order to compute the TEE of the FCHI, we use the Kitaev-Preskill construction~\cite{PhysRevLett.96.110404} extended to 3D systems~\cite{PhysRevB.84.195120}. As sketched in Fig.~\ref{fig:SketchKitaevPreskill3D}(b), the system is divided into four regions $\mathcal{A}$, $\mathcal{B}$, $\mathcal{C}$ and $\mathcal{D}$, which are translation invariant in the $z$ direction and whose cross sections with the $xy$-plane form the pattern required for the usual 2D Kitaev-Preskill cut. The EE of these regions and their unions can be collected into the linear combination
	\begin{equation}\label{TEE}
	-\gamma = S^{(2)}_{\mathcal{A}\mathcal{B}\mathcal{C}} - S^{(2)}_{\mathcal{A}\mathcal{B}} - S^{(2)}_{\mathcal{B}\mathcal{C}} - S^{(2)}_{\mathcal{A}\mathcal{C}} + S^{(2)}_{\mathcal{A}} + S^{(2)}_{\mathcal{B}} + S^{(2)}_{\mathcal{C}}
	\end{equation}
	which cancels all contributions from the virtual surfaces and hinges. The remaining quantity, denoted $\gamma$, could contain two contributions $\gamma = \gamma_{3D} + N_z \times \gamma_{2D}$. The constant $\gamma_{3D}$ is the 3D TEE~\cite{PhysRevB.84.195120}. $\gamma_{2D}N_z$ would occur for layered constructions of 2D topological orders perpendicular to the $z$ direction with 2D TEE $\gamma_{2D}$~\cite{PhysRevB.84.195120} or in some fractonic systems~\cite{doi:10.1146/annurev-conmatphys-031218-013604,doi:10.1142/S0217751X20300033}.
	
	We have computed $\gamma$ for the FCHI on the 3-torus in large-scale variational MC computations. For the geometry sketched in Fig.~\ref{fig:SketchKitaevPreskill3D}(b), we were able to study the FCHI with $3\times 3 \times 2$ unit cells, where we found $\gamma = - 0.08 \pm 0.04$, and with $3\times 3 \times 3$ unit cells, where we found $\gamma = -0.06 \pm 0.11$. In both cases, the subsystem $\mathcal{A}$ is of size $1\times 2 \times N_z$ unit cells, and the subsystems $\mathcal{B}$ and $\mathcal{C}$ are of size $1\times 1 \times N_z$ unit cells~\footnote{We recall that the number of physical sites for these subsystems is $2\times 4 \times N_z$ and $2\times 2 \times N_z$, respectively}. Because of the intrinsic anisotropy of the FCHI, also considered a second geometry obtained by rotating the subsystems in Fig.~\ref{fig:SketchKitaevPreskill3D}(b) along the $y$ axis such that they are translation invariant in the $x$ direction, while leaving the insulator unchanged. Here, we computed $\gamma$ for a system of $2\times 3 \times 5$ unit cells~\footnote{We recall that the number of physical sites in this case is $4 \times 6 \times 5$. Thus despite $N_x = 2$, the PBC along $x$ are non-trivial. Here, the subsystem $\mathcal{A}$ is of size $2\times 2 \times 2$ unit cells, and the subsystems $\mathcal{B}$ and $\mathcal{C}$ are of size $2\times 1 \times 2$ unit cells} and found $\gamma = -0.009 \pm 0.102$. All these values are consistent with $\gamma = 0$ (up to small finite-size effects for $3\times 3 \times 2$) \emph{irrespective} of the orientation of the cut. We stress that $\gamma$ is several orders of magnitude smaller that any of the EE appearing in Eq.~\eqref{TEE}, excluding the existence of both a non-vanishing 3D TEE $\gamma_{3D}$ and a non-zero $\gamma_{2D}$.
	
	Since we have not been able to find any clear signature of a true non-trivial bulk topology, we now probe the nature of the gapped surfaces perpendicular to the $x$ direction~\footnote{We could alternatively choose those perpendicular to the $y$ direction.}. Since the vertical hinges host fractionalized one-dimensional modes like those of an FCI, we may speculate that the vertical surfaces host some non-trivial TO~\cite{PhysRevLett.124.046801}. To characterize it, we compute $\gamma$ according to Eq.~\eqref{TEE} for the geometry obtained by rotating the subsystems in Fig.~\ref{fig:SketchKitaevPreskill3D}(b) as described above, OBC in the $x$ direction and PBC in the $y$ and $z$ directions. We have performed this computation for a system with $2\times 3 \times 5$ unit cells and found $\gamma = 0.31 \pm 0.20$~\footnote{Here, the statistical error bar is a conservative estimate, in particular much larger than the remaining fluctuations in the mean of the MC computations~\reftosuppmat{sec:AppendicMCTechnical}. We stress that our conservative error bar confirms a non-trivial TEE smaller than $\log \sqrt{2}$ per surface.}. Since the same computation with PBC in $x$ yields a vanishing result for $\gamma$ as discussed above, this non-zero value is due entirely to the two surfaces at $x = 0$ and $x = N_x -1$ and confirms that the vertical surfaces host a non-trivial 2D TO. The value for $\gamma$ is consistent with $\ln \sqrt{2}$, the TEE of a \emph{single} 2D FCI in the Laughlin $1/2$ phase, which would imply that each of the surface TOs has a TEE of $(\ln \sqrt{2})/2$.
	
	We mention that we have also studied the degeneracy of the four ansatz states for the FCHI in this geometry, which are generated by changing the boundary conditions for the underlying CHI in the two periodic directions. We have found very similar behavior to the the full-PBC case discussed above, namely two larger eigenvalues but no clear evidence of a reduction of the bulk degeneracy in the thermodynamic limit~\reftosuppmat{sec:AppendixFCHI}.

	\section{Discussion and conclusion}
	
	We have studied a model wave function for a 3D chiral hinge insulator with strong interactions at fractional band filling. By studying the EE and spin fluctuations in an open geometry, we showed that the hinges host fractional gapless modes which have the same characterization as the edge modes of an FCI in the Laughlin $1 / 2$ phase. We have also studied the system's topology through the topological degeneracy and the TEE. While the results for the topological degeneracy remain inconclusive due to the small number of numerically accessible system sizes, our results point to the absence of a bulk TEE. However, we found clear signatures of a non-trivial 2D topological order on the vertical surfaces. Interestingly, the TEE contribution per surface is consistent with $(\ln \sqrt{2})/2$, in other words half of the TEE of an FCI, which cannot be described using a quantum dimension~\footnote{A TEE of $(\ln \sqrt{2})/2$ would correspond to a total quantum dimension $2^{1/4}$, whereas any non-trivial total quantum dimension has to be larger or equal to $\sqrt{2}$.}. This suggests a non-trivial relation between the surface topology and the hinge modes~\cite{PhysRevLett.124.046801}. In this letter, we have restricted our analysis to the gapped surfaces and their gapless edges. It would be highly interesting but very numerically challenging to consider the top and bottom surfaces, which host single Dirac cones in the non-interacting CHI~\cite{InPrep}. Their fate in the interacting system is yet unknown and beyond the scope of the present work, but it should be the focus of further study.

	\section{Acknowledgment}
	
	We thank Benoit Estienne for enlightening discussions. A.Hackenbroich and N.S. acknowledge support by the European Research Council (ERC) under the European Union's Horizon 2020 research and innovation programme through the ERC Starting Grant WASCOSYS (No. 636201) and the ERC Consolidator Grant SEQUAM (No. 863476), and by the Deutsche Forschungsgemeinschaft (DFG) under Germany's Excellence Strategy (EXC-2111 – 390814868). A.Hackenbroich and N.R. were supported by Grant No. ANR-17-CE30-0013-01. N.R. was partially supported by NSF through the Princeton University’s Materials Research Science and Engineering Center DMR-2011750B. A.Hudomal acknowledges funding provided by the Institute of Physics Belgrade, through the grant by the Ministry of Education, Science, and Technological Development of the Republic of Serbia. Part of the numerical simulations were performed on the PARADOX-IV supercomputing facility at the Scientific Computing Laboratory, National Center of Excellence for the Study of Complex Systems, Institute of Physics Belgrade. B.A.B. was supported by the DOE Grant No. DE-SC0016239, the Schmidt Fund for Innovative Research, Simons Investigator Grant No. 404513, the Packard Foundation, the NSF-EAGER No. DMR 1643312, NSF-MRSEC No. DMR-1420541 and DMR-2011750, ONR No. N00014-20-1-2303, Gordon and Betty Moore Foundation through Grant GBMF8685 towards the Princeton theory program, BSF Israel US foundation No. 2018226, and the Princeton Global Network Funds.
	
	\bibliography{HigherOrder}		
	
\clearpage
\onecolumngrid

\begin{center}
	\textbf{\large Supplemental Material for "Fractional Chiral Hinge Insulator"}
\end{center}
\appendix

\setcounter{figure}{0}
\setcounter{table}{0}
\renewcommand{\thefigure}{S\arabic{figure}}
\renewcommand{\thetable}{S\arabic{table}}

\tableofcontents


\listoffigures
	
\addtocontents{toc}{\protect\setcounter{tocdepth}{3}}
\addtocontents{lot}{\protect\setcounter{lotdepth}{3}}	
\addtocontents{lof}{\protect\setcounter{lofdepth}{3}}

\section*{Introduction to the Appendices}\label{App:AppendixOverview_appendix}	

In this work, we have built and studied a model state for a three-dimensional (3D) fractional chiral hinge insulator (FCHI). In the following, we provide additional information on our numerical methods,  their benchmark on a two-dimensional (2D) fractional Chern insulator and the non-interacting 3D chiral hinge insulator, and additional results for the topological degeneracy of the 3D model. 

We begin in Appendix~\ref{sec:AppendixMC} by introducing the Monte Carlo (MC) observables that we used for the computations of entanglement entropy and the overlap matrix whose results are discussed in the main text. In Appendix~\ref{sec:AppendixFCI}, we proceed to give a detailed account of the benchmark of our MC algorithm and the entanglement observables from the main text on a 2D fractional Chern insulator. This includes a characterization of the edge modes from variational MC simulations, and the computation of the topological entanglement entropy and topological degeneracy. In Appendix~\ref{sec:CHI}, we complement the characterization of the fractional hinge modes of the FCHI in the main text by a similar analysis for the underlying non-interacting chiral hinge insulator, confirming that its hinge modes are of the same nature as the edge modes of a non-interacting Chern insulator. In Appendix~\ref{sec:AppendixFCHI}, we give additional details about the computation of the topological overlap matrix for the FCHI whose eigenvalue scaling for isotropic system sizes is discussed in the main text. We also provide results for the scaling of the overlap matrix for highly anisotropic systems that are much larger in the $z$ direction than in the $x$ and $y$ directions, and for geometries with open boundary conditions in the $x$ direction and periodic boundary conditions in the $y$ and $z$ directions. Finally, in Appendix~\ref{sec:AppendicMCTechnical} we provide technical data about our MC computations, including the computational cost.

\section{Monte Carlo simulations\label{sec:AppendixMC}}

In this appendix, we provide a quick overview of the Monte Carlo (MC) procedure that was used to derive the results in the main text. The technical details of the MC simulation will be discussed in Appendix~\ref{sec:AppendicMCTechnical}.

\subsection{Entanglement entropy\label{sec:MCEntanglementEntropy}}

For a quantum state $\ket{\psi}$, the Renyi entropy of order $n$ \wrt a subsystem $\mathcal{A}$ is given by 
\begin{equation}
S^{(n)}_{\mathcal{A}} = \frac{1}{1 - n} \ln \left( \tr_{\mathcal{A}} \left[ \rho_{\mathcal{A}} ^n \right]\right).
\end{equation}
Here, $\rho_{\mathcal{A}} = \tr_{\mathcal{B}} \left[ \ket{\psi} \bra{\psi}\right]$ is the reduced density matrix of the subsystem $\mathcal{A}$ obtained by tracing over the degrees of freedom in its complement $\mathcal{B}$. We use the notation $\tr_{\mathcal{A}}$ for the trace over the degrees of freedom in region $\mathcal{A}$. In the limit $n \rightarrow 1$, $S^{(n)}$ corresponds to the von Neumann entropy $S_{\mathcal{A}} = - \tr_{\mathcal{A}} \left[ \rho_{\mathcal{A}} \ln \left( \rho_{\mathcal{A}} \right)\right]$. Since the von Neumann entropy is hard to evaluate numerically, we focus on the second Renyi entropy $S^{(2)}$, which can be computed using the replica trick with the $\swap$ operator technique~\cite{PhysRevLett.104.157201}.

To that end, we consider two identical copies of the system, which together are in the quantum state $\ket{\psi} \otimes \ket{\psi}$. The $\swap$ operator acts by exchanging the degrees of freedom within the subsystem $\mathcal{A}$ between the two copies, while leaving the degrees of freedom in $\mathcal{B}$ unchanged. Specifically, let $\{\ket{v}\}$ be an orthonormal basis of (a single copy of) the system, whose elements $\ket{v} = \ket{v_{\mathcal{A}}, v_{\mathcal{B}}}$ factorize as a tensor product of states describing only the subsystem $\mathcal{A}$ and its complement $\mathcal{B}$, respectively. Expressed in this basis, the $\swap$ operator acts on the two copies as 
\begin{equation}
\swap \left(\ket{v} \otimes \ket{v'}\right) = \swap \left(\ket{v_{\mathcal{A}}, v_{\mathcal{B}}} \otimes \ket{v'_{\mathcal{A}}, v'_{\mathcal{B}}} \right) = \ket{v'_{\mathcal{A}}, v_{\mathcal{B}}} \otimes \ket{v_{\mathcal{A}}, v'_{\mathcal{B}}}.
\end{equation}
It can be shown that the second Renyi entropy is related to the expectation value $\myexp{\swap}$ as~\cite{PhysRevLett.104.157201}
\begin{equation}
e^{ - S_{\mathcal{A}}^{(2)}} = \tr_{\mathcal{A}}\left[ \rho_{\mathcal{A}} ^2\right] = \myexp{\swap} = \frac{\bra{\psi \otimes \psi} \swap \ket{\psi \otimes \psi}}{\myexp{\psi \otimes \psi |\psi \otimes \psi  }}.
\end{equation} 
The expectation value $\myexp{\swap}$ can be computed using MC simulations on the double-copy system. Computations of the entanglement entropy with this method have been successfully performed in the context of topological phases for systems such as spin liquids~\cite{PhysRevB.84.075128, PhysRevB.87.161113}, Laughlin states~\cite{PhysRevB.84.075128} and non-Fermi-liquids~\cite{PhysRevLett.114.206402}. 

Since the Renyi entropy of a quantum ground state obeys an area law and the expectation value $\myexp{\swap}$ measured in MC simulations is given by $e^{ - S_{\mathcal{A}}^{(2)}}$, the value of $\myexp{\swap}$ decays exponentially with $|\partial \mathcal{A}|$. Therefore, for larger subsystems the convergence of $\myexp{\swap}$ quickly becomes extremely slow. This can partially be mitigated by a sign trick~\cite{PhysRevLett.107.067202} for non-positive wave functions, which allows to separately evaluate the contributions to $\myexp{\swap}$ from the amplitude and the phase of the wave function. To that end, we write
\begin{equation}
\myexp{\swap} = \myexp{\swapAmp} \times \myexp{\swapPhase},
\end{equation}
where $\myexp{\swapAmp}$ and $\myexp{\swapPhase}$ can be measured in separate MC simulations with faster convergence. Denoting by $\psi(v) \equiv \psi(v_{\mathcal{A}}, v_{\mathcal{B}})\equiv \myexp{v | \psi}$ the coefficient of the quantum state $\ket{\psi}$ \wrt the basis state $\ket{v}$, the two expectation values are given by~\cite{PhysRevLett.107.067202}
\begin{subequations}\label{SWAPAmpPhase}
\begin{gather}\label{SWAPAmp}
\myexp{\swapAmp} = \sum_{v, v'} \rho (v, v')_{\text{amp}} \times f (v, v')_{\text{amp}},\\
\myexp{\swapPhase} = \sum_{v, v'} \rho (v, v')_{\text{phase}} \times f (v, v')_{\text{phase}},
\end{gather}
where
\begin{gather}
	\rho (v, v')_{\text{amp}} = \frac{|\psi(v)|^2}{\myexp{\psi| \psi}} \frac{|\psi(v')|^2}{\myexp{\psi| \psi}} ,\\
	f (v, v')_{\text{amp}} = \left| f (v, v') \right|,\\
	\rho (v, v')_{\text{phase}} = \frac{|\psi (v'_{\mathcal{A}}, v_{\mathcal{B}}) \psi (v_{\mathcal{A}}, v'_{\mathcal{B}})\psi(v_{\mathcal{A}}, v_{\mathcal{B}}) \psi(v'_{\mathcal{A}}, v'_{\mathcal{B}})|} { \sum_{v, v'}|\psi (v'_{\mathcal{A}}, v_{\mathcal{B}}) \psi (v_{\mathcal{A}}, v'_{\mathcal{B}})\psi(v_{\mathcal{A}}, v_{\mathcal{B}}) \psi(v'_{\mathcal{A}}, v'_{\mathcal{B}})|},\\
	f (v, v')_{\text{phase}} = e^{i\arg[f (v, v')]},
\end{gather}
\end{subequations}
and we used the shorthand
\begin{equation}
f (v, v') \equiv \frac{\psi (v'_{\mathcal{A}}, v_{\mathcal{B}}) \psi (v_{\mathcal{A}}, v'_{\mathcal{B}})}{\psi(v_{\mathcal{A}}, v_{\mathcal{B}}) \psi(v'_{\mathcal{A}}, v'_{\mathcal{B}})}
\end{equation}
split into its norm $|f (v, v')|$ and its phase, $\arg[f (v, v')]$. 
The expectation values $\myexp{\swapAmp}$ and $\myexp{\swapPhase}$ can be evaluated using MC simulations with the probability densities $\rho (v, v')_{\text{amp}}$ and $\rho (v, v')_{\text{phase}}$ and estimators $f (v, v')_{\text{amp}}$ and $f (v, v')_{\text{phase}}$, respectively. We note that for the entropy computations in the FCHI presented in Fig.~\ref{fig:EEChiralHingeTi} of the main text, the convergence of the $\myexp{\swapPhase}$ observable is much faster than that of the $\myexp{\swapAmp}$ observable. Similar observations have been made for RVB states, where this was linked to an approximate Marshall sign rule~\cite{PhysRevB.88.125135}.

\subsection{Wave function overlap\label{sec:AppendixMCOverlap}}

In Appendix~\ref{sec:AppendixFCHI}, we will compute the overlap matrix of the different ansatz states for the FCHI on the 3D torus in order to check if the system has a topological degeneracy. The overlap matrix element $\mathcal{O}_{\psi_1,\psi_2}$ between two a priori unnormalized ansatz states $\ket{\psi_1}$ and $\ket{\psi_2}$ is given by
\begin{equation}
\mathcal{O}_{\psi_1,\psi_2} = \frac{\myexp{\psi_1 | \psi_2}}{\sqrt{\myexp{\psi_1 | \psi_1}} \sqrt{\myexp{\psi_2 | \psi_2}}}.
\end{equation}

In order to compute this overlap with MC, we make use of the identity
\begin{equation}
\mathcal{O}_{\psi_1,\psi_2} = \sum_{v} \frac{\psi_1(v)^* \,\psi_2(v)}{\sqrt{\myexp{\psi_1 | \psi_1}} \sqrt{\myexp{\psi_2 | \psi_2}}} = \sqrt{\mathcal{O}_{\psi_1,\psi_2}^{1, abs} \times \mathcal{O}_{\psi_1,\psi_2}^{2, abs}} \times \mathcal{O}_{\psi_1,\psi_2}^{ phase},
\end{equation}
where
\begin{subequations}\label{DefOverlapObsMC}
	\begin{gather}
	\mathcal{O}_{\psi_1,\psi_2}^{1, abs} = \sum_{v} \rho(v)^{1, abs} \times f(v)^{1, abs},\\
	\mathcal{O}_{\psi_1,\psi_2}^{2, abs} = \sum_{v} \rho(v)^{2, abs} \times f(v)^{2, abs},\\
	\mathcal{O}_{\psi_1,\psi_2}^{ phase} = \sum_{v} \rho(v)^{phase} \times f(v)^{phase}.
	\end{gather}
\end{subequations}
	and
\begin{subequations}	
	\begin{gather}
		\rho(v)^{1, abs} = \frac{\left| \psi_1(v)\right|^2}{\myexp{\psi_1 | \psi_1}} ,\\
		f(v)^{1, abs} = \frac{|\psi_2(v)|}{|\psi_1(v)|} ,\\
		\rho(v)^{2, abs} = \frac{\left| \psi_2(v)\right|^2}{\myexp{\psi_2 | \psi_2}} ,\\
		f(v)^{2, abs} = \frac{|\psi_1(v)|}{|\psi_2(v)|},\\
		\rho(v)^{phase} = 	\frac{ |\psi_2(v)||\psi_1(v)| }{\sum_{v} |\psi_2(v)||\psi_1(v)|},\\
		f(v)^{phase} = e^{i(\arg[\psi_2(v)] - \arg[\psi_1(v)])}.		
	\end{gather}
\end{subequations}
In Eq.~\eqref{DefOverlapObsMC}, the sum is over all configurations $v$. As before, $\arg[\psi_i(v)]$ refers to the phase of the wave function $\psi_i(v) = \myexp{v | \psi_i}$ for $i = 1,2$. The quantities $\mathcal{O}_{\psi_1,\psi_2}^{1, abs}$, $\mathcal{O}_{\psi_1,\psi_2}^{2, abs}$ and $\mathcal{O}_{\psi_1,\psi_2}^{ phase}$ can be evaluated in separate MC computations with weights $\rho(v)^{1, abs}$, $\rho(v)^{2, abs}$, $\rho(v)^{phase}$ and measurement estimators $f(v)^{1, abs}$, $f(v)^{2, abs}$ and $f(v)^{phase}$, respectively. It is also possible to express the overlap matrix element as the product of three expectation values that can be sampled in MC computations with the same weight~\cite{PhysRevB.94.045125}.

In order to compare the physical validity of different unnormalized ansatz states, it can be useful to compare their norms. To that end, we note that the ratio of the norms of the ansatz states $\ket{\psi_1}$, $\ket{\psi_2}$ is given by
\begin{equation}
\frac{\myexp{\psi_1 | \psi_1}}{\myexp{\psi_2 | \psi_2}} = \frac{\mathcal{O}_{\psi_1,\psi_2}^{2, abs}}{\mathcal{O}_{\psi_1,\psi_2}^{1, abs}},
\end{equation} 
and is therefore a by-product of the MC computations for the overlap matrix.

\section{Fractional Chern insulator\label{sec:AppendixFCI}}

In this section, we test the concepts developed in the main text and benchmark our tools by considering a fractional Chern insulator (FCI) in two dimensions. The latter is obtained as the Gutzwiller projection of two copies of a simple two-band Chern insulator (CI) described in Ref.~\onlinecite{PhysRevB.84.075128}. We begin by briefly presenting the non-interacting CI model in Sec.~\ref{sec:CI}. In Sec.~\ref{sec:FCI_EdgeModes}, we then benchmark the methods used in the main text for the hinge mode characterization on the edge modes of the FCI. Subsequently, we study the stability of the FCI under the addition of a staggered chemical potential in Sec~\ref{sec:FCI_TEE}, before concluding with a study of the topological degeneracy in Sec.~\ref{sec:FCI_TopDeg}.

\subsection{Chern insulator\label{sec:CI}}

\begin{figure*}[t]	
	\begin{tikzpicture}
	\node at (0, 0) {\includegraphics[width = 0.15\linewidth] {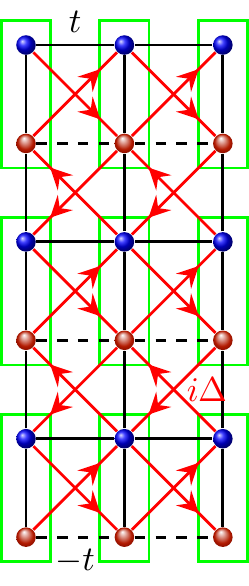}};
	\node at (-2, 2.5) {(a)};
	\node at (6, 0) {\includegraphics[width = 0.3\linewidth] {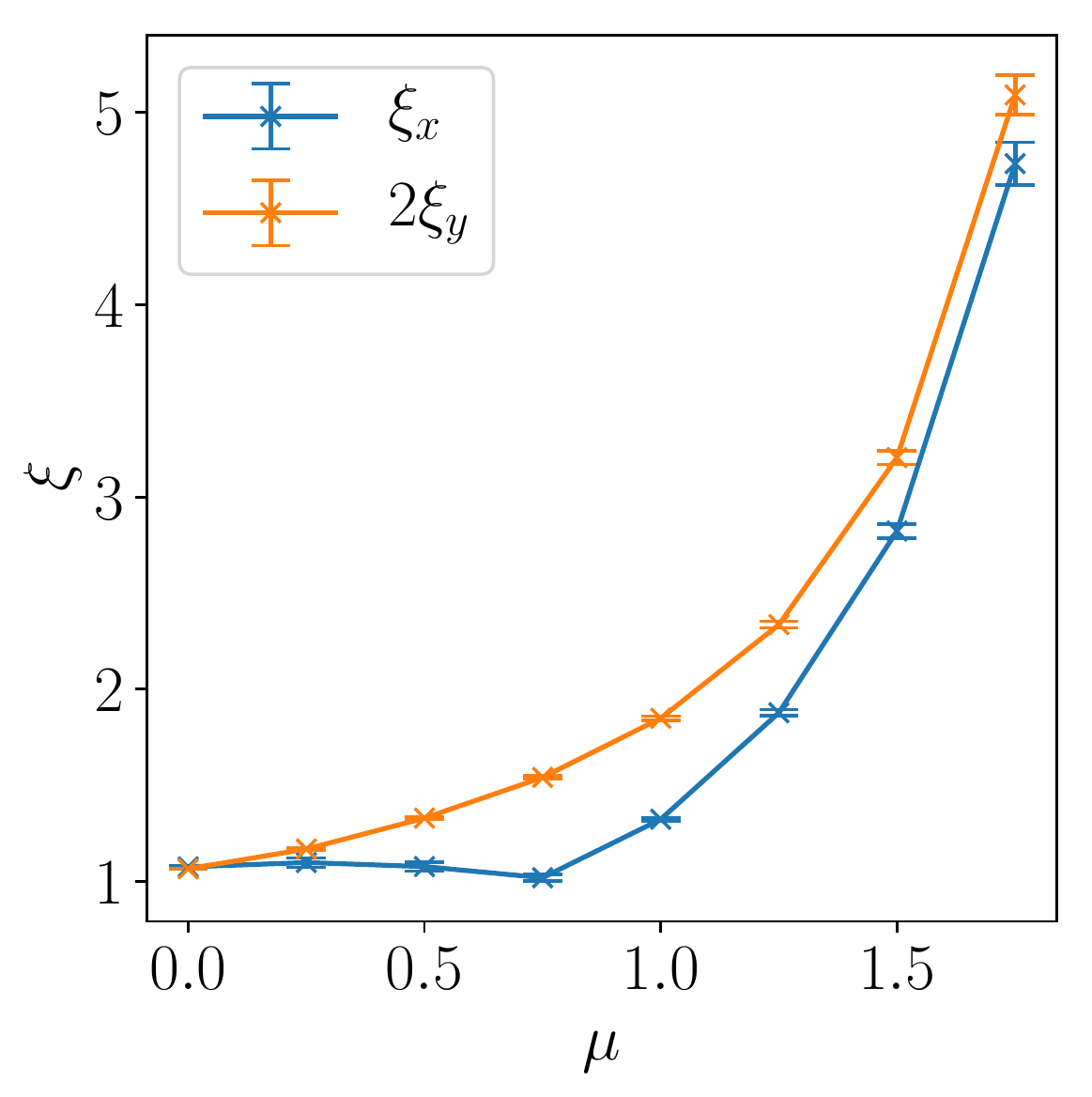}};
	\node at (3, 2.5) {(b)};
	\end{tikzpicture}
	\caption[Tight-binding models for CI]{(a) Microscopic model for the Chern insulator of Ref.~\onlinecite{PhysRevB.84.075128} defined on a square lattice with two sublattices, $A$ in blue and $B$ in red. The nearest-neighbour hopping $t$ ($-t$ for dashed lines) drawn in black is real, whereas the next-nearest neighbour hopping $i\Delta$ in the direction of the red arrows is purely imaginary. (b) Correlation lengths $\xi_x$ and $\xi_y$ (in units of the unit cell) for the two-point correlator in the ground state of the perturbed CI with staggered chemical potential $\mu$. The correlation lengths diverge at the critical value $\mu_c = 2$. Note that due to the anisotropy of the unit cell, $\xi_x$ and $2\xi_y$ are comparable. 
	\label{fig:SketchCRealSpaceCI}}
\end{figure*}

We consider the two-band Chern insulator (CI) of Ref.~\onlinecite{PhysRevB.84.075128} at half filling whose valence band has a non-zero Chern number $\mathcal{C} = 1$. The corresponding tight-binding Hamiltonian sketched in Fig.~\ref{fig:SketchCRealSpaceCI}(a) is defined on a square lattice with $N_x$ unit cells in the horizontal direction and $N_y$ unit cells in the vertical direction, where each unit cell consists of an $A$ and a $B$ site. It is characterized by a real next-nearest neighbour hopping $t$ and a purely imaginary next-to-nearest neighbour hopping $i\Delta$. We will also consider an additional staggered chemical potential with $+\mu$ on $A$ sites and $-\mu$ on $B$ sites. The Bloch Hamiltonian of the CI is given by
\begin{equation}\label{CI_Hamiltonian}
H \left(k_x, k_y \right)= \left[ 2 \Delta \sin(k_x) \left(\cos(k_y) -1\right) + t \left(\cos(k_y) + 1\right)\right] \sigma_x + \left[ \left(2 \Delta \sin(k_x) + t\right) \sin(k_y)\right] \sigma_y + \left[ \mu + 2t \cos(k_x)\right] \sigma_z,
\end{equation}
where $\sigma_x$, $\sigma_y$ and $\sigma_z$ are the Pauli matrices. The Bloch bands have energy $\pm \epsilon \left(k_x, k_y \right) $, with
\begin{equation}\label{CI_BandEnergy}
\epsilon \left(k_x, k_y \right) = \sqrt{\left[ 2 \Delta \sin(k_x) \left(\cos(k_y) -1\right) + t \left(\cos(k_y) + 1\right)\right] ^2 + \left[ \left(2 \Delta \sin(k_x) + t\right) \sin(k_y)\right] ^ 2+\left[ \mu + 2t \cos(k_x)\right]^2}.
\end{equation}
For $\mu = 0$, the single-particle gap is maximal when the hopping parameters are $t = 1$ and $\Delta = 1 / 2$. In the following, we therefore always choose $t = 1$ and $\Delta = 1 / 2$.

Increasing $\mu$ away from $0$ leads to a trivialization of the model (i.e., when the two bands have a zero Chern number) for $\mu$ larger than a critical value $\mu_c$ where the band gap closes. For $t = 1$ and $\Delta = 1 / 2$, the single-particle gap closes at $\mu_c = 2$. As expected, the correlation length of the CI ground state at half filling diverges as $\mu$ approaches $\mu_c$, but stays reasonably small for values $\mu \leq 1$ (see Fig.\ref{fig:SketchCRealSpaceCI}(b)).

\subsubsection{Twisted boundary conditions}\label{sec:TwistedBC}

On the torus, we can consider the CI with twisted boundary conditions determined by phases $\Phi_x$ and $\Phi_y$ that a particle should pick up on a loop parallel to the $x$ and $y$ axes, respectively. Here, we choose to implement the twisted boundary conditions in the tight-binding model in a translation-invariant way by multiplying all hopping terms in the positive $x$ and $y$ directions with phases $\lambda_x$ and $\lambda_y$, respectively, where
\begin{subequations}\label{FluxesCI}
	\begin{gather}
	\lambda_x = e^{i \frac{\Phi_x}{N_x}},\\
	\lambda_y = e^{i \frac{\Phi_y}{2N_y}}.						
	\end{gather}
\end{subequations}
Correspondingly, hopping terms with a component in the negative $x$ and $y$ directions are multiplied with the complex conjugate phases $\lambda_x^*$ and $\lambda_y^*$.

\subsubsection{Particle-hole symmetry}

The unperturbed CI model with zero staggered chemical potential $\mu = 0$ possesses a unitary particle-hole (PH) symmetry, which relates states with different boundary conditions on the torus. The PH conjugation acts on the creation operators as
\begin{subequations}\label{DressedPH}
	\begin{gather}
	c_{A, (x,y)} \mapsto c_{A, (x,y)}^{\dagger},\\
	c_{B, (x,y)} \mapsto - c_{B, (x,y)}^{\dagger},
	\end{gather}
\end{subequations}
where $x \in \{0, \dotsc, N_x -1\}$ and $y \in \{0, \dotsc, N_y -1\}$ are the unit cell coordinates. It is straightforward to verify that on the torus, this symmetry maps the CI Hamiltonian with twist phase factors $(\lambda_x, \lambda_y)$ to the CI Hamiltonian with modified phase factors $(\lambda_x', \lambda_y')$, given by
 \begin{equation}
\left(\lambda_x', \lambda_y' \right) = \left( - \lambda_x^*, \lambda_y^* \right).
\end{equation}
Using Eq.~\eqref{FluxesCI}, this implies that the non-interacting ground state with $(\Phi_x, \Phi_y)$ is mapped to the ground state with
\begin{equation}\label{FluxMapping}
\left(\Phi_x', \Phi_y'\right) = \left( - \Phi_x + N_x\pi, - \Phi_y\right) 
\end{equation}
Note that for $N_x$ odd, the change in $\Phi_x$ is an odd multiple of $\pi$, such that the symmetry relates states with periodic and anti-periodic boundary conditions in the horizontal direction.

\subsection{FCI edge modes\label{sec:FCI_EdgeModes}}

The FCI obtained as the Gutzwiller projection of two copies of the CI at half filling lies in the same universality class as the bosonic Laughlin state at filling $\nu = 1/2$. In particular, its chiral gapless edge modes are described by the chiral CFT $\mathfrak{su}(2)_1$. Two key characteristic quantities of this CFT, the central charge $c = 1$ and the Luttinger parameter $K = 1/2$, can be extracted numerically from the scaling of the entanglement entropy (EE) and spin fluctuations in a suitable geometry, respectively~\cite{crepel2019model, crepel2019microscopic,estienne2019entanglement}. Here, we reproduce these known results with our MC setup using a geometry which can easily be generalized to the 3D setting of the main text. This serves both as a benchmark for our numerical tools, and as a validation of the geometry used here and in the main text.

\begin{figure}[t]	
	\begin{tikzpicture}
	\node at (0, 0) {\includegraphics[width = 0.4\linewidth] {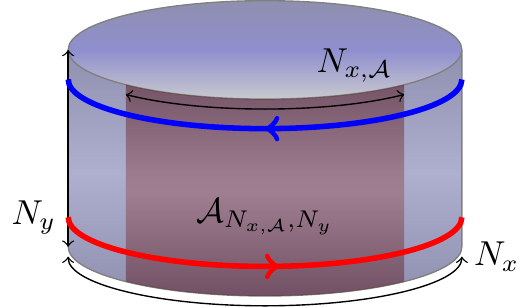}};
	\node at (-4.3, 2.5) {(a)};
	\node at (9, 0) {\includegraphics[width = 0.5\linewidth] {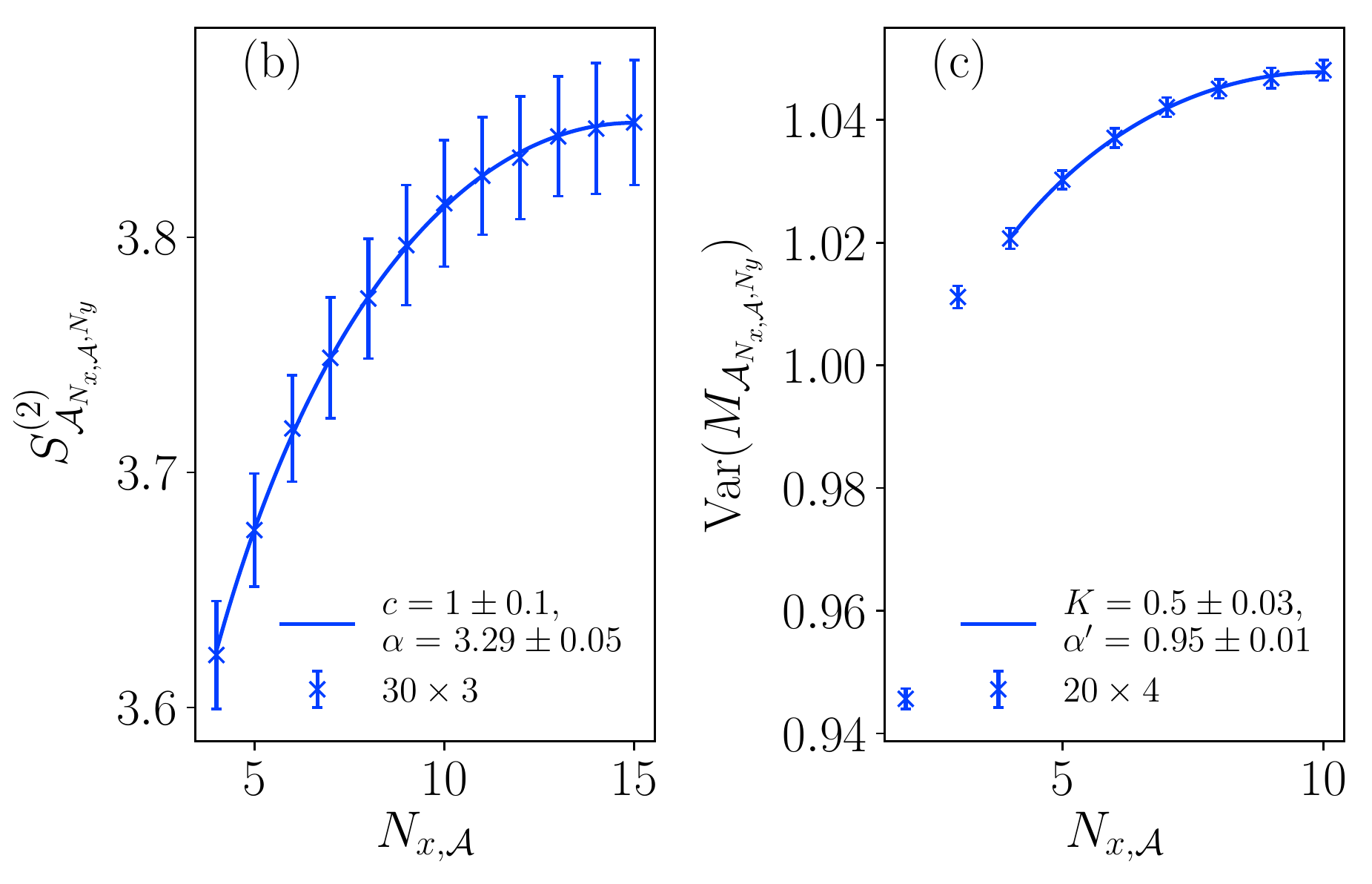}};
	\end{tikzpicture}
	\caption[Edge mode characterization]{(a) Geometry used for the extraction of the edge physics: Cylinder with periodic boundary conditions and $N_x$ unit cells in the horizontal direction, and open boundaries and $N_y$ sites in the vertical direction. The rectangular subsystem $\mathcal{A}_{N_{x, \mathcal{A}}, N_{y}}$ consists of $N_{x, \mathcal{A}}$ unit cells in the horizontal direction and $N_{y}$ sites in the vertical direction. (b), (c)~MC results for the EE and spin fluctuations characterizing the edge modes extracted from the geometry in (a).\label{fig:EdgeModes}}
\end{figure}

We consider the FCI in the ``ribbon'' geometry sketched in Fig.~\ref{fig:EdgeModes}(a). The system is defined on a cylinder with periodic boundary conditions and $N_x$ unit cells in the $x$-direction, and open boundaries and $N_y$ unit cells in the $y$-direction. We consider the EE and spin fluctuations \wrt a series of subsystems $\mathcal{A}_{N_{x, \mathcal{A}}, N_{y}}$ which have $N_{x, \mathcal{A}} \in \{1, N_x/2\}$ unit cells in the horizontal direction and span the full height of the cylinder in the vertical direction. The $\mathcal{A}_{N_{x, \mathcal{A}}, N_{y}}$ cuts break the translation invariance in the horizontal direction and introduce virtual boundaries perpendicular to the physical edge states (marked in red and blue). Hence, the EE and spin fluctuations \wrt $\mathcal{A}_{N_{x, \mathcal{A}}, N_{y}}$ will contain contributions from both chiral edge modes, in addition to the area law and corner contribution which do not depend on $N_{x, \mathcal{A}}$.

Concretely, the second Renyi entropy $S^{(2)}$ is expected to scale with $N_{x, \mathcal{A}}$ as
\begin{equation}\label{EERibbon}
S^{(2)}_{\mathcal{A}_{N_{x, \mathcal{A}}, N_{y}}} (N_{x, \mathcal{A}}) = \alpha_{2D} + 2 \times S^{(2)}_{\text{crit}} (N_{x, \mathcal{A}}; N_x).
\end{equation}	
Here, $\alpha_{2D}$ contains the area law contributions proportional to $N_y$ from the virtual cuts, as well as the corner contributions or constant corrections. It is therefore a constant independent of $N_{x, \mathcal{A}}$. In Eq.~\eqref{EERibbon}, 	
\begin{equation}\label{EE1DCriticalApp}
S^{(2)}_{\text{crit}} (N_{x, \mathcal{A}}; N_x) = \frac{c}{8} \ln \left[ \frac{N_x}{\pi} \sin \left( \frac{\pi N_{x, \mathcal{A}}}{N_x}\right)\right]
\end{equation}
is the second Renyi entropy of a periodic one-dimensional chiral critical mode with central charge $c$ and total system size $N_x$ restricted to a single interval of length $N_{x, \mathcal{A}}$~\cite{Calabrese_2009}. The factor of $2$ in Eq.~\eqref{EERibbon} takes into account the two edge modes, which each contribute equally to the EE.

In the main text, we also considered the variance of the number $M_{\mathcal{A}}$ of spin up particles in the region $\mathcal{A}$, given by
\begin{equation}\label{eq:SpinFluctuations}
 \var(M_{\mathcal{A}})\equiv\langle M_{\mathcal{A}}^2\rangle-\langle M_{\mathcal{A}}\rangle^2.
\end{equation}
For the $\mathcal{A}_{N_{x, \mathcal{A}}, N_y}$ cut used here, the variance is
expected to scale as~\cite{estienne2019entanglement}
\begin{equation}\label{FitFunctionParticleNumberVariance}
\var(M_{\mathcal{A}_{N_{x, \mathcal{A}}, N_y}}) = \alpha'_{2D} + \frac{K} {\pi ^ 2}  \ln \left[ \frac{N_x}{\pi} \sin \frac{\pi  N_{x, \mathcal{A}}} {N_x} \right] ,
\end{equation}
where $K$ is the Luttinger parameter of the CFT describing the edge modes, and $\alpha'_{2D}$ is a constant independent of $N_{x, \mathcal{A}}$ subsuming the corner and area law contributions to the spin fluctuations.

The scaling of the EE and spin fluctuations for the FCI as computed from MC, fit to the predictions of Eqs.~\eqref{EERibbon} and \eqref{FitFunctionParticleNumberVariance}, are shown in Fig.~\ref{fig:EdgeModes}(b) and (c). In both cases, the logarithmic contributions from the edge states are clearly visible. The fit values of the central charge $c = 1.0\pm 0.1$ and the Luttinger parameter $K = 0.50 \pm 0.03$ are very close to the expected values $c = 1$ and $K = 1/2$, respectively.

\subsection{Topological entanglement entropy\label{sec:FCI_TEE}}

We now explore the stability of the FCI under the perturbation by a staggered chemical potential $\mu$ as defined in Eq.~\eqref{CI_Hamiltonian}. To this end, we compute the topological entanglement entropy (TEE)~\cite{PhysRevLett.96.110404} denoted $\gamma$. For the FCI obtained as the Gutzwiller projection of two copies of the non-interacting CI, the TEE is expected to be $\gamma = \ln (2) / 2 \approx  0.347$.

In order to test our MC calculations, we first extract the TEE using exact diagonalization (ED) and the Kitaev-Preskill scheme~\cite{PhysRevLett.96.110404}. Here by ED we mean that the ground state at half filling of the CI is obtained directly by ED, providing the ground state decomposition onto the real space many-body basis. With such a decomposition, we can easily perform the Gutzwiller projection and exactly compute any EE, while the EE calculation using MC is considerably more complex, as described in Appendix~\ref{sec:MCEntanglementEntropy}. For the system size we study here, ED is fast enough and provides exact results without any errors. It can therefore be used as a reference for MC calculations.

\begin{figure*}[t]
	\begin{tikzpicture}
	\node at (0, 0) {\includegraphics[width = 0.25\linewidth] {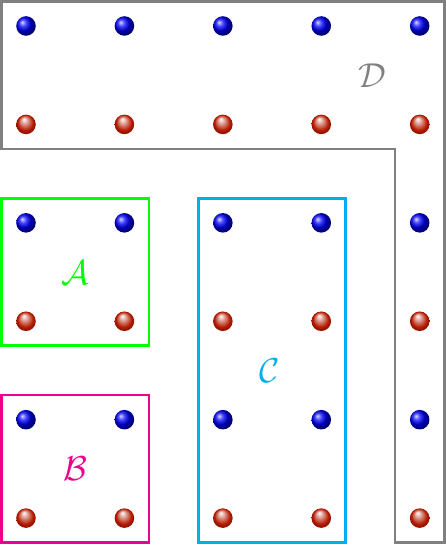}};
	\node at (-2.8, 3) {(a)};
	\node at (8, 0) {\includegraphics[width = 0.45\linewidth] {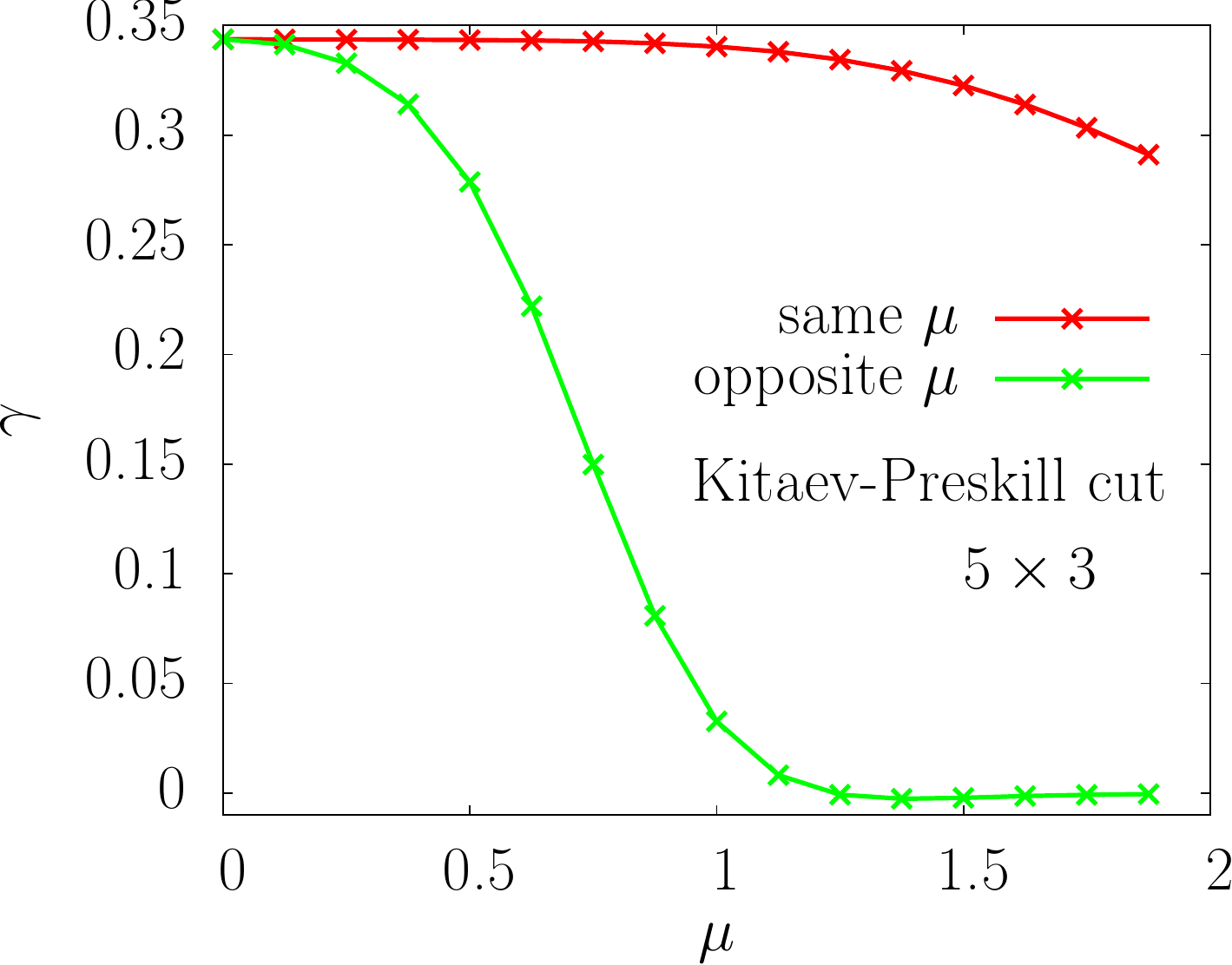}};
	\node at (3.5, 3) {(b)};
	\end{tikzpicture}	
	\caption[Topological entanglement entropy for FCI with staggered chemical potential]{(a) Sketch of the Kitaev-Preskill cut for a system of $5\times3$ unit cells. (b) Topological entanglement entropy for the FCI as a function of staggered chemical potential $\mu$. The results were obtained for the system size $5\times3$ using exact diagonalization and the Kitaev-Preskill scheme shown in (a). 
	\label{fig:FCI_KitaevPreskill}}
\end{figure*}

We study a system of size $3\times5$ unit cells with periodic boundary conditions in both directions and a cut shown in Fig.~\ref{fig:FCI_KitaevPreskill}(a). The system is divided into four regions labeled $\mathcal{A}$, $\mathcal{B}$, $\mathcal{C}$ and $\mathcal{D}$ (see Fig.~\ref{fig:FCI_KitaevPreskill}(a)). The linear combination of the entanglement entropies of these regions and their unions 
\begin{equation}
 S_{\mathcal{A}\mathcal{B}\mathcal{C}}- S_{\mathcal{A}\mathcal{B}}- S_{\mathcal{B}\mathcal{C}}- S_{\mathcal{A}\mathcal{C}}+ S_{\mathcal{A}}+ S_{\mathcal{B}}+ S_{\mathcal{C}}
\end{equation}
suppresses all unwanted contributions (such as the area law or corner contributions) and only the TEE term remains.

The TEE for the FCI as a function of staggered chemical potential $\mu$ is presented in Fig.~\ref{fig:FCI_KitaevPreskill}(b). We consider two different cases, same $\mu$ in both non-interacting CI copies (i.e., before projection) and mixed $\mu$ ($+\mu$ in one copy and $-\mu$ in the other). As can be seen from Fig.~\ref{fig:SketchCRealSpaceCI}(b), for $\mu \leq 1$ the correlation length of the CI ground state is shorter than two lattice spacings of the underlying square lattice. We may therefore expect that in this case the subsystem sizes considered here are large enough compared to the correlation length that our computation gives a finite size TEE close to the result in the thermodynamic limit. Indeed, for $\mu=0$, the numerically obtained value is $\gamma \approx 0.344$, which is close to the predicted value of $0.347$. The agreement is expected to be even better in larger systems which are not accessible in our ED calculations. In the case of same $\mu$ (red line), the TEE stays approximately constant for $\mu\leq1$ and then it starts to deviate due to increasing correlation length. In contrast, the TEE for opposite $\mu$ (green line) immediately decreases from the expected value and drops to zero shortly after $\mu=1$. We can conclude that the FCI is only stable to addition of a staggered chemical potential with the same sign in both CI copies, while it is destroyed by a staggered chemical potential with opposite signs.

Finally, we repeat the calculation for system size $5\times3$ and $\mu=0$ using MC and the Kitaev-Preskill scheme. The TEE obtained in this way is $\gamma=0.34\pm0.03$, which is in good agreement with the ED result and the theoretical prediction.

\subsection{Topological degeneracy\label{sec:FCI_TopDeg}}

In Appendix~\ref{sec:AppendixFCHI} we will discuss the topological degeneracy of the fractional chiral hinge insulator with periodic boundary conditions in all directions. For sake of comparison, here we compute the topological degeneracy of the FCI on the torus. The topological degeneracy is given by the number of linearly independent states generated by Gutzwiller projection of the non-interacting CI with twisted boundary conditions characterized by the phases $\Phi_x$ and $\Phi_y$ as introduced in Appendix~\ref{sec:TwistedBC} above. Denoting by $\ket{\psi^{ (\Phi_x, \Phi_y)}_{\sigma}}$ the ground state of the CI model with twisted boundary conditions with spin $\sigma \in \{\uparrow, \downarrow\}$, we compute the rank of the overlap matrix
\begin{equation}
\left(P_G\left[\bra{\psi^{(\Phi_x, \Phi_y)}_{\uparrow}} \otimes \bra{\psi^{(\Phi_x, \Phi_y)}_{\downarrow}}\right] P_G\left[\ket{\psi^{(\Phi_x', \Phi_y')}_{\uparrow}} \otimes \ket{\psi^{(\Phi_x', \Phi_y')}_{\downarrow}}\right]\right)_{(\Phi_x, \Phi_y), (\Phi_x', \Phi_y')}.
\end{equation}
For simplicity, we only consider phases $\Phi_x, \Phi_y,\Phi_x', \Phi_y' \in \{0, \pi\}$. This choice ensures that the Gutzwiller projected state has periodic boundary conditions, while the underlying electronic system has periodic or anti-periodic boundary conditions.

\subsubsection{Unperturbed FCI ($\mu=0$)\label{FCI_TD_mu0}}

We use ED, as discussed in Appendix~\ref{sec:FCI_TEE}, in order to obtain the rank of the overlap matrix for the FCI in different system sizes. In contrast to MC where we need to perform an independent calculation for each matrix element of the overlap matrix, as explained in Appendix~\ref{sec:AppendixMCOverlap}, in the case of ED it is enough to compute the four Gutzwiller projected states $P_G\left[\ket{\psi^{(\Phi_x, \Phi_y)}_{\uparrow}} \otimes \ket{\psi^{(\Phi_x, \Phi_y)}_{\downarrow}}\right]$ for different combinations of phases $(\Phi_x, \Phi_y)$. The full overlap matrix is then straightforwardly obtained by simply computing the scalar products of these states.

The expected topological degeneracy for the FCI is 2. For system sizes with odd $N_x$, there are indeed two pairs of linearly dependent states, thus an exact twofold degeneracy. This exact degeneracy is due to the dressed particle-hole symmetry of the CI model which relates states with different boundary conditions $(\Phi_x, \Phi_y)$ and $(\Phi_x', \Phi_y')$ in accordance with Eq.~\eqref{FluxMapping}. However, when $N_x$ is even all states are linearly independent and the overlap matrix rank is equal to 4. In this case, each of the four states is mapped to itself by the PH conjugation. It is important to note that for even $N_x$ there are two eigenvalues approximately equal to 2 and two very small eigenvalues (see Fig.~\ref{fig:FCI_OverlapEV_SameMu}(b) and (c) for $\mu=0$). As will be shown in the next subsection, the two largest eigenvalues are approaching 2 with increasing system size, while the two smallest eigenvalues are decreasing towards zero. This points to the conclusion that the overlap matrix rank will be equal to 2 in the thermodynamic limit, as expected.

We note that we first normalize the states obtained by the Gutzwiller projection of two copies of the CI ground state and then generate the overlap matrix and compute its rank. This may lead to numerical inaccuracy in larger systems where the weight of the Gutzwiller projected state is small, as numerical errors might be significantly amplified by normalization. It is also possible to generate the overlap matrix using unnormalized states. However, the trace of the overlap matrix will no longer be equal to its dimension in that case.

\subsubsection{Staggered chemical potential\label{sec:FCI_TD_mu}}

We also consider the effect of staggered chemical potential $\mu$ on the eigenvalues and the rank of the overlap matrix. 
The key idea is to move away from the dressed PH symmetry at $\mu=0$ that enforces exactly two linearly independent states for odd $N_x$, knowing the nature of the FCI should be unchanged.
We again consider the FCI with PBC in both directions and perform ED.

\begin{figure*}[t]
	\begin{tikzpicture}
	\node at (-6, 0) {\includegraphics[width = 0.32\linewidth] {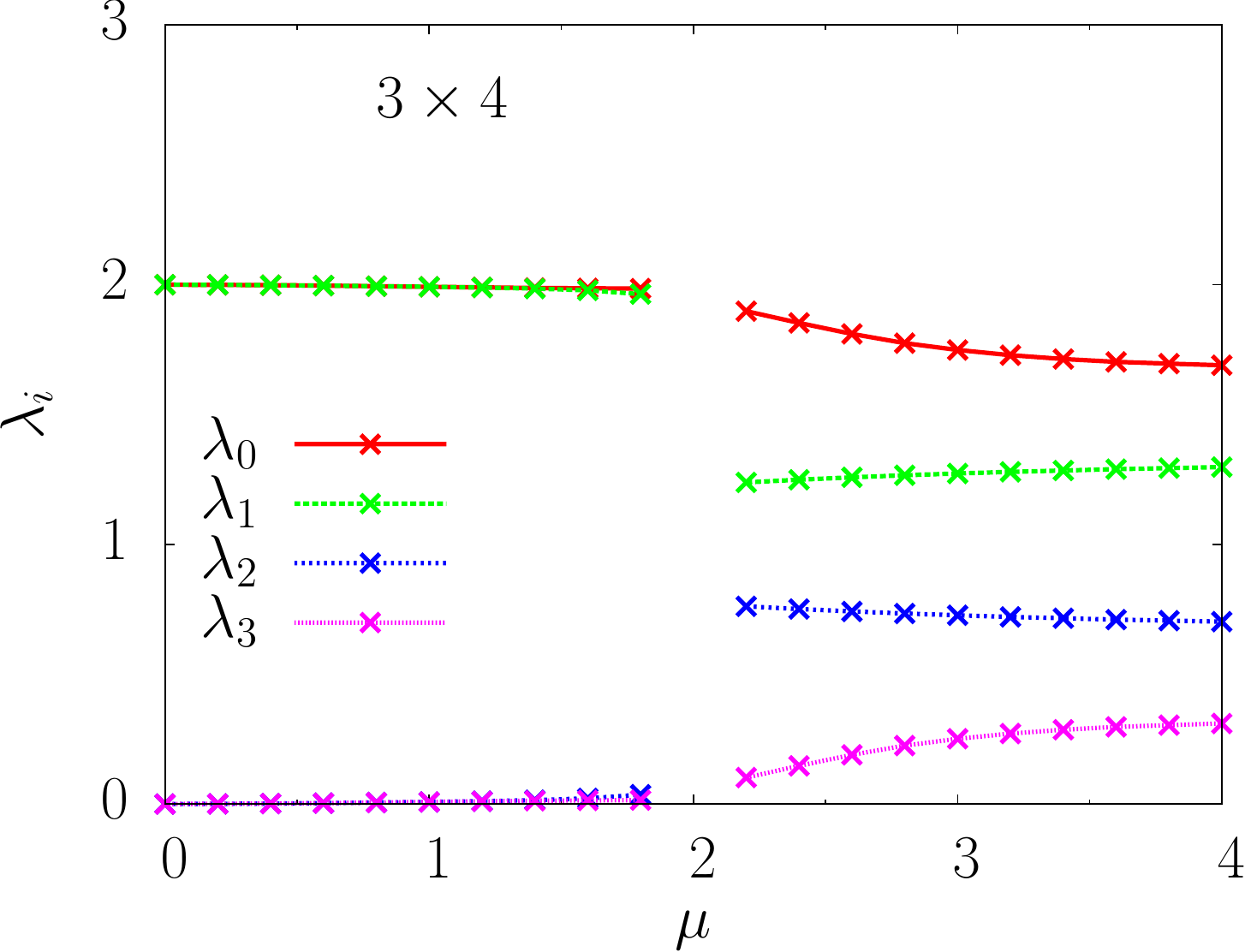}};
	\node at (-8.8, 2.1) {(a)};
	\node at (0, 0) {\includegraphics[width = 0.32\linewidth] {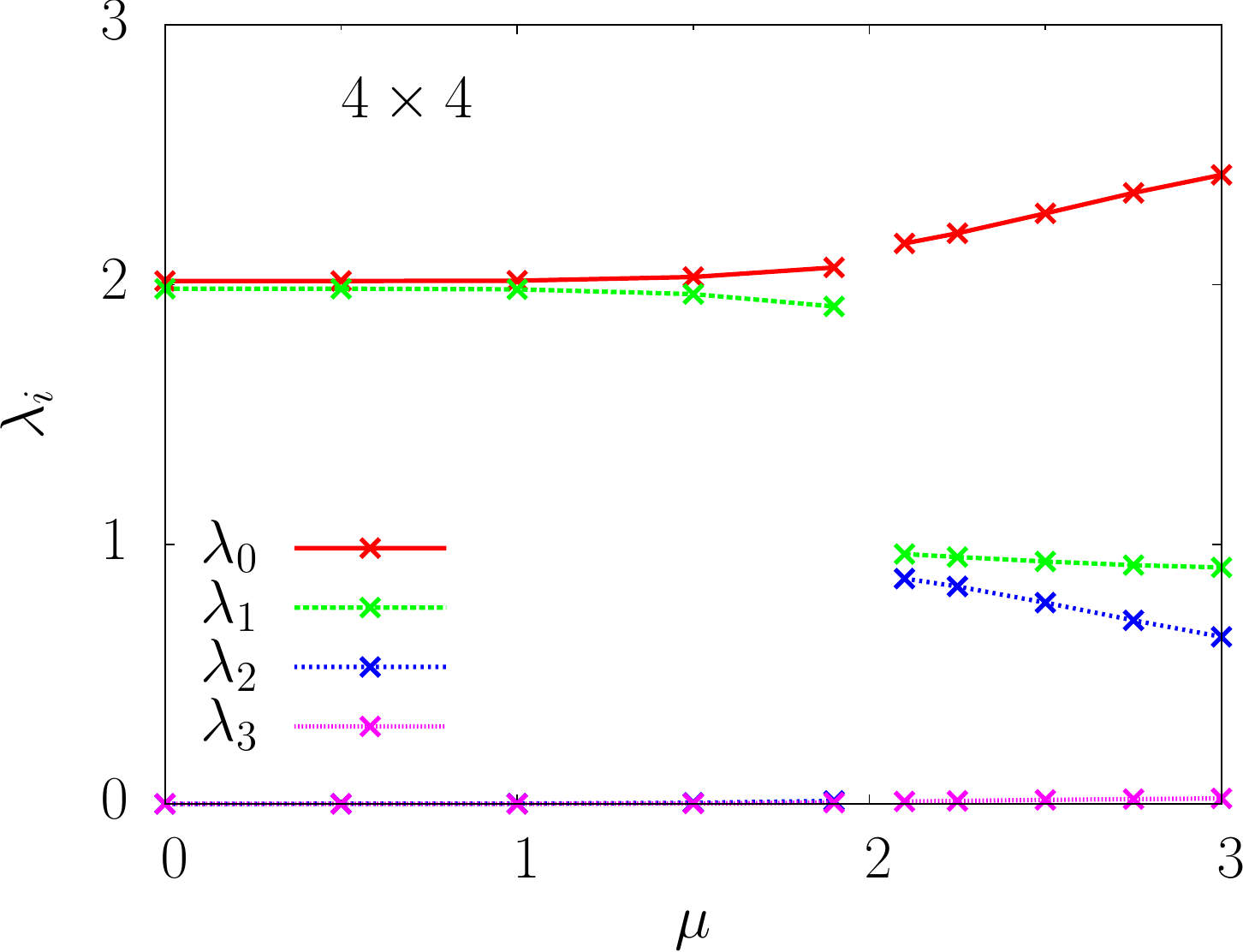}};
	\node at (-2.8, 2.1) {(b)};
	\node at (6, 0) {\includegraphics[width = 0.32\linewidth] {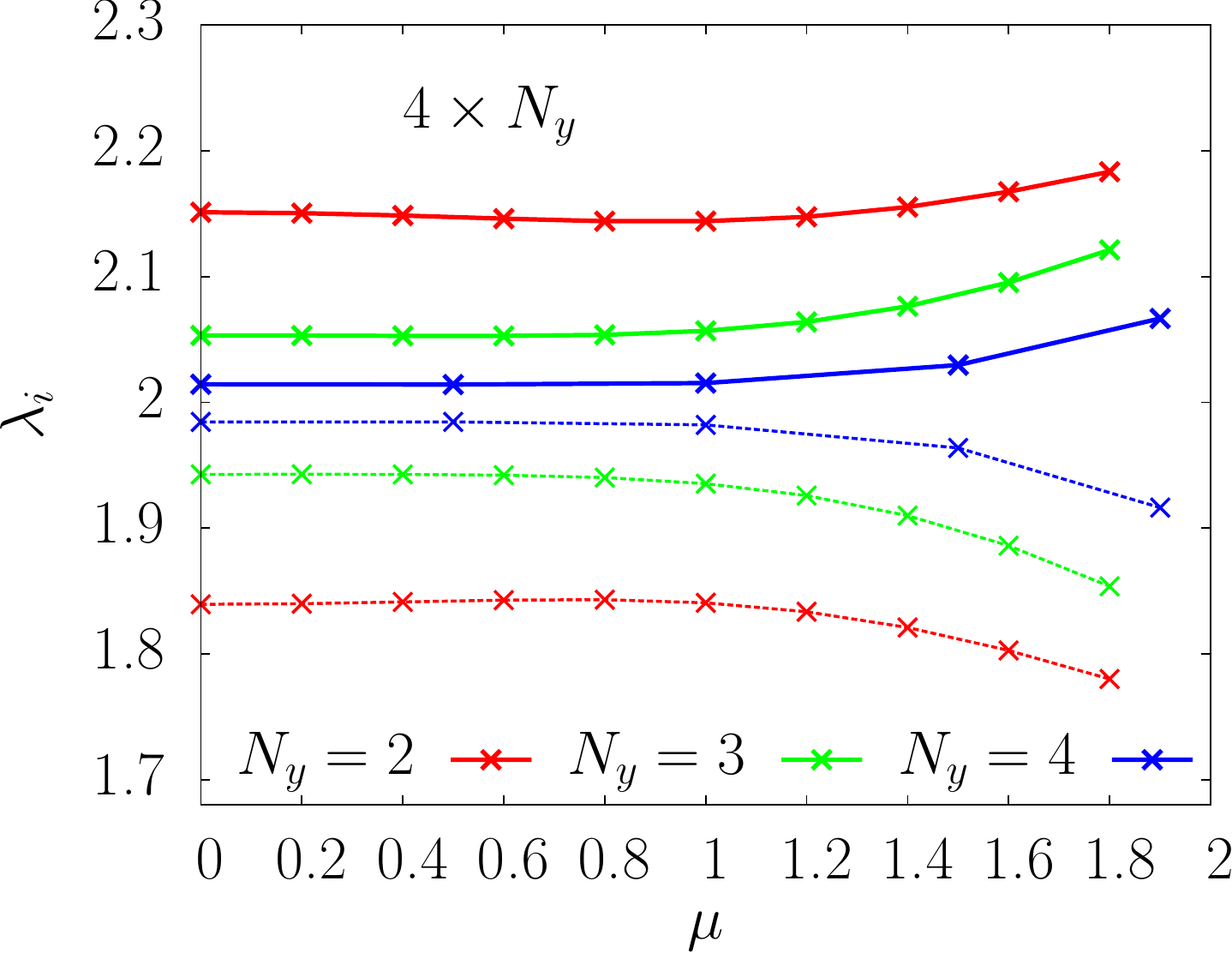}};
	\node at (3.2, 2.1) {(c)};
	\end{tikzpicture}	
	\caption[FCI Overlap matrix eigenvalues]{Eigenvalues of the overlap matrix computed from ED for the FCI as a function of the staggered chemical potential $\mu$ for different system sizes. In all cases, the chemical potential is the same in both copies of the CI wave function underlying the FCI. The size of the torus is in (a) $3 \times 4$ and in (b) $4 \times 4$. In both cases, the overlap matrix for $\mu < \mu_c = 2$ has two very small eigenvalues and two eigenvalues close to 2, indicating a topological degeneracy of 2 as expected for the FCI. The phase transition at $\mu_c=2$ is clearly visible in the discontinuity of some eigenvalues, with an \textit{a priori} unclear phase for $\mu > \mu_c$ (see text). In (c), comparison of the scaling of the two largest overlap matrix eigenvalues for three different system sizes $4 \times N_y$ with $N_y \in \{2, 3, 4\}$ in the FCI phase with $\mu < \mu_c = 2$. The eigenvalues approach the value $2$ with increasing $N_y$, implying due to the conserved trace of the overlap matrix that the remaining two eigenvalues (not shown) approach the value $0$.
	\label{fig:FCI_OverlapEV_SameMu}}
\end{figure*}

The eigenvalues $\lambda_i$ of the overlap matrix as a function of $\mu$ for two different system sizes can be observed in Fig.~\ref{fig:FCI_OverlapEV_SameMu}(a) and (b). In both cases, some of the eigenvalues clearly have a discontinuity at $\mu_c=2$. As previously discussed in Sec.~\ref{sec:CI}, a staggered potential larger than the critical value $\mu_c=2$ trivializes the CI model. The number of nonzero eigenvalues (rank of the overlap matrix) in the region $\mu>2$ is 4. However, the same is formally true even in the FCI phase, except for odd $N_x$ at $\mu=0$ where there is an exact degeneracy due to the dressed PH symmetry. As already discussed in the previous subsection for even $N_x$ and $\mu=0$, the main difference is that in the topological phase there are two very small eigenvalues and two eigenvalues of the overlap matrix close to 2. Fig.~\ref{fig:FCI_OverlapEV_SameMu}(c) shows the scaling of the two largest overlap matrix eigenvalues for system sizes with even $N_x$. The two largest eigenvalues are approaching the value 2 with increasing system size. At the same time, the other two eigenvalues are decreasing towards zero (not shown here), as the sum of all eigenvalues must be exactly 4. These results suggest that the rank of of the overlap matrix in the FCI phase is 2 in the thermodynamic limit. 

We note that the Gutzwiller projection for $\mu>2$ might not be meaningful. In the limit of large $\mu$, all the particles are located at B sites in the CI ground state. The Gutzwiller projection removes double occupancies, meaning that the projected state in the high-$\mu$ region will consist only of particle fluctuations. These fluctuations are then squared during the overlap matrix calculation, which may further lead to numerical instability. We therefore cannot infer the nature of the phase beyond $\mu_c=2$. In contrast, the Gutzwiller projection of two CI ground state copies with opposite $\mu$ is well defined, as there are no doubly occupied sites in the $\mu\to\infty$ limit: one copy has the electrons sitting on A sites while the other copy has its electrons sitting on B sites. The Gutzwiller projected state is this a perfect Mott insulator. However, the TEE calculations in Appendix~\ref{sec:FCI_TEE} have shown that the FCI is not stable to the addition of opposite staggered chemical potential in the two CI copies. Although the system is in that case clearly in the trivial phase for $\mu>2$, the nature of the phase for a Gutzwiller projection of two states with opposite $\mu$ is unknown for $\mu<2$.  
	
\section{Chiral hinge HOTI\label{sec:CHI}}

	In this appendix, we revisit the non-interacting chiral hinge HOTI~\cite{schindler2018higher} whose Gutzwiller projection leads to the FCHI wave function studied in the main text. We begin in Sec.~\ref{sec:HamiltonianCHI} with a quick review of the tight-binding Hamiltonian. In Sec.~\ref{sec:HingeStatesCHI}, we then characterize the non-interacting hinge states via their EE and particle number fluctuations analogously to the analysis of the FCHI hinge modes in the main text. This allows us to test our MC methods on a similar system where a direct, non-interacting, calculation is available.
	
	\subsection{Tight-binding model\label{sec:HamiltonianCHI}}

	The second-order chiral hinge TI was introduced using a band structure which can be realized both in a tight-binding model for spin-1/2 electrons as well as an optical lattice set-up with spinless fermions~\cite{schindler2018higher}. Here, we study a variant of the latter realization with an additional staggered chemical potential of magnitude $\mu$. The model is described by a local Hamiltonian for spinless fermions on the cubic lattice with four sites per unit cell lying in the $xy$ plane, labeled $1$ to $4$ as sketched in Fig.~\ref{fig:SketchCHI}(a) of the main text. In this plane, sites in the same unit cell are connected by a nearest-neighbour hopping $M$, whereas sites in adjacent unit cells are connected by a nearest-neighbour hopping $\Delta_1$. Additionally, there is a $\pi$-flux through each plaquette in the $xy$-plane. In the $z$-direction, adjacent unit cells are connected by a real next-nearest neighbour hopping $- \Delta_2/ 2$, and a purely imaginary nearest neighbour hopping with value $-i\Delta_2 / 2$. In addition, we consider a staggered chemical potential with $+\mu$ on sites 1 and 2 and $-\mu$ on sites 3 and 4. All in all, the Bloch Hamiltonian of this model is	
	\begin{multline}\label{BlochHamiltonianChiralHingeTIStaggeredMu}
H (k_x, k_y, k_z)= \left[M + \Delta_1 \cos(k_x) - \Delta_2 \cos(k_z) \right] \tau_x \sigma_0 
+ \left[M + \Delta_1 \cos(k_y) - \Delta_2 \cos(k_z) \right] \left( - \tau_y \sigma_y \right) \\
+ \Delta_1 \sin(k_x) \left( - \tau_y \sigma_z \right) + \Delta_1 \sin(k_y) \left( - \tau_y \sigma_x \right) 
+ \left[\mu -\Delta_2 \sin(k_z) \right]\left( \tau_z \sigma_0 \right),
\end{multline}
	where $\tau_x, \tau_y, \tau_z, \sigma_x, \sigma_y, \sigma_z$ are the Pauli matrices and $\sigma_0$ the identity matrix acting on the sublattice degree of freedom. The valence and conduction bands of the CHI model are doubly degenerate with energies $\pm \epsilon$, where
	\begin{equation}
	\epsilon^2 = \left[M + \Delta_1 \cos(k_x) - \Delta_2 \cos(k_z) \right]^2 + \left[M + \Delta_1 \cos(k_y) - \Delta_2 \cos(k_z) \right] ^2 + \Delta_1^2 \left[ \sin(k_x)^2 + \sin(k_y)^2 \right] + (\mu - \Delta_2 \sin(k_z))^2.
	\end{equation}
	We consider the model at half filling $\nu = 1/2$.
	
	Note that the CHI model in the topological phase can be seen as a trivial-to-topological dipole pumping interpolation of the topological quadrupole model of Refs.~\cite{benalcazar2017quantizedScience, benalcazar2017quantized}. Indeed, for each fixed $k_z$, the Bloch Hamiltonian of Eq.~\eqref{BlochHamiltonianChiralHingeTIStaggeredMu} for $\mu = 0$ defines an instance of the two-dimensional quadrupole model with a staggered chemical potential. With the notation of Eq.~(VI.55) of Ref.~\cite{benalcazar2017quantized}, the Hamiltonian parameters $\delta(k_z)$, $\lambda(k_z)$ and $\gamma(k_z)$ of this two-dimensional model at fixed $k_z$ are related to the parameters $M$, $\Delta_1$ and $\Delta_2$ of the three-dimensional model as
	\begin{subequations}
		\begin{gather}
		\delta(k_z) = - \Delta_2 \sin(k_z),\\
		\lambda = \Delta_1,\\
		\gamma (k_z)= M - \Delta_2 \cos(k_z).
		\end{gather}
	\end{subequations}  
	For $M = \Delta_1 = \Delta_2 = 1$ and $\mu = 0$, the CHI model is in its topological phase: Indeed, the parameters of the quadrupole model at $k_z = 0$ are $(\delta, \lambda, \gamma) = (0, 1, 0)$, so we get the obstructed atomic limit phase of the quadrupole model. For $k_z = \pi$, the parameters are $(\delta, \lambda, \gamma) = (0, 1, 2)$, so the model is in the trivial phase of the quadrupole model. Therefore, the CHI model realizes a trivial-to-topological interpolation of the topological quadrupole model, under which the corner states of the quadrupole model map to the hinge states of the CHI model.
	
\begin{figure*}[t]
	\begin{tikzpicture}

	\node at (0, 0) {\includegraphics[width = \linewidth] {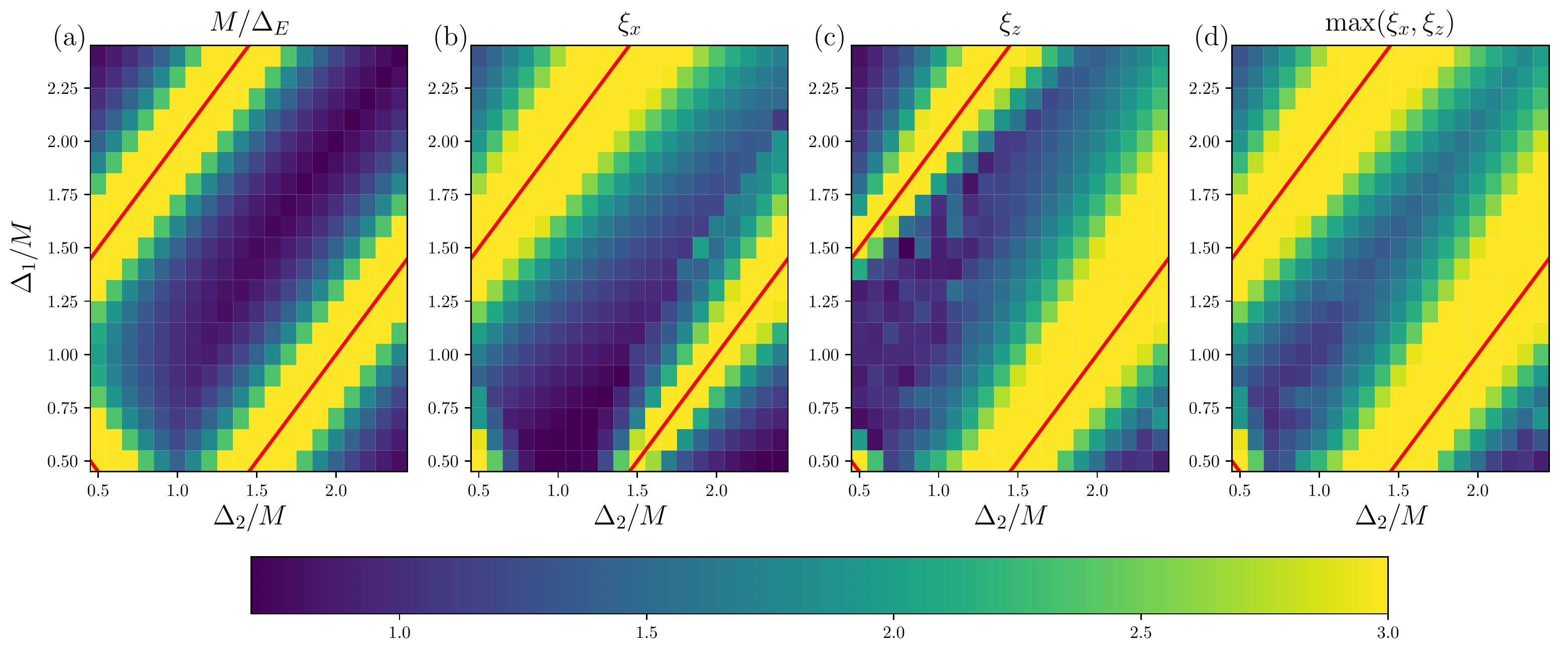}};

	\end{tikzpicture}	
	\caption[Correlation length CHI]{Inverse gap and correlation lengths of the CHI model of Eq.~\eqref{BlochHamiltonianChiralHingeTIStaggeredMu} for different Hamiltonian couplings $\Delta_1 / M$ and $\Delta_2 / M$ when $\mu = 0$. The model in in its topological phase in the region around the point $M = \Delta_1 = \Delta_2 = 1$, bordered by phase transitions when $\Delta_2 / M = \Delta_1 / M \pm 1$ and $\Delta_2 / M = - \Delta_1 / M \pm 1$ (marked by red lines). In (a), inverse of the minimal direct gap $\Delta_E$ in units of $M$, \ie $M / \Delta_E$. In (b) and (c), horizontal and vertical correlation lengths $\xi_x ( = \xi_y)$ and $\xi_z$, respectively, in units of unit cells and computed from the two-point correlation function. In (d), the largest correlation length $\max(\xi_x, \xi_z)$, which attains its minimal value close to the point $\Delta_1 / M = \Delta_2 / M = 1$. \label{fig:CorrelationLength}}
\end{figure*}

	At $\mu = 0$, the topological phase of the CHI model around the point $M = \Delta_1 = \Delta_2 = 1$ is bordered by phase transitions lines when $\Delta_2 / M = \Delta_1 / M \pm 1$ and $\Delta_2 / M = - \Delta_1 / M \pm 1$. At these parameter values, the minimal direct gap 
	\begin{equation}
	\Delta_E \equiv \min_{k_x, k_y, k_z} \epsilon(k_x, k_y, k_z)
	\end{equation}
	of the CHI Hamiltonian of Eq.~\eqref{BlochHamiltonianChiralHingeTIStaggeredMu} closes. This can be seen in Fig.~\ref{fig:CorrelationLength}(a), where we show the inverse of the minimal direct gap in units of $M$. The horizontal and vertical correlation lengths $\xi_x$ (which is equal to $\xi_y$) and $\xi_z$ shown in Fig.~\ref{fig:CorrelationLength}(b) and (c), respectively, diverge around the phase transition lines. Note that due to the intrinsic anisotropy of the CHI model, the correlation lengths in the vertical $z$ direction and the horizontal $x$ and $y$ directions do not need to be identical. The minimal direct gap reaches its maximal value $\sqrt{2}$ for parameters $\Delta_2 / M = \Delta_1 / M \geq 2$. However, the smallest value of the larger of the two correlation lengths, $\max(\xi_x, \xi_z)$ shown in Fig.~\ref{fig:CorrelationLength}(d), is obtained for parameter values close to $\Delta_2 / M = \Delta_1 / M = 1$. Since we want to minimize the finite-size effects, we therefore always choose $M = \Delta_1 = \Delta_2 = 1$.
	
\begin{figure*}[t]
	\begin{tikzpicture}
	\node at (9.5, 0) {\includegraphics[width = 0.4\linewidth] {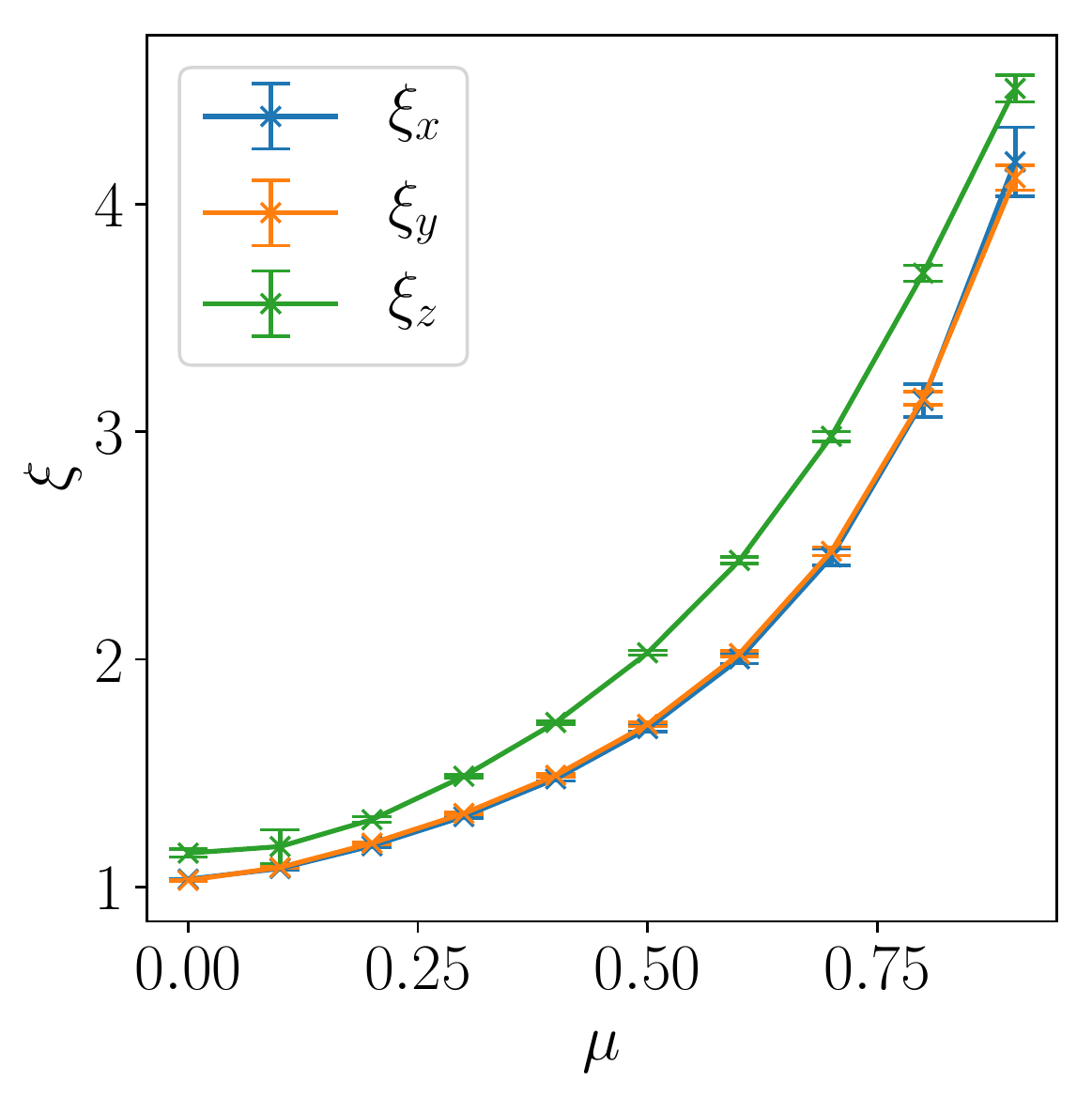}};

	\end{tikzpicture}	
	\caption[Correlation length CHI finite $\mu$]{Correlation lengths $\xi_x$ , $\xi_y$ and $\xi_z$ (in units of the unit cell) for the two-point correlator of the CHI as a function of staggered chemical potential $\mu$ for $M = \Delta_1 = \Delta_2 = 1$.\label{fig:CorrelationLength_Mu}}
\end{figure*}

	Increasing $\mu$ away from zero at half filling leads to a trivialization of the model for $\mu$ larger than a critical value $\mu_c$. For $M = \Delta_1 = \Delta_2 = 1$, the single-particle gap closes at $\mu_c = 1$. As expected, the correlation length of the CHI ground state diverges as $\mu$ approaches $\mu_c$, but stays reasonably small for values $\mu \leq 1/2$ (see Fig.\ref{fig:CorrelationLength_Mu}).
	
	The CHI model is invariant under the product $\cfourz \mathcal{T}$ of the fourfold rotation $\cfourz$ and time reversal $\mathcal{T}$~\cite{schindler2018higher}. This is the symmetry which protects the higher-order topological phase. In addition, the model is invariant under the product $I \mathcal{T}$ with the inversion $I$~\cite{schindler2018higher}.

Similarly to the 2D case discussed in Appendix~\ref{sec:TwistedBC}, on the 3D torus we can also consider the CHI with twisted boundary conditions set by phases $\Phi_x$, $\Phi_y$ and $\Phi_z$ that a particle should pick up on a loop in $x$, $y$ and $z$ direction, respectively. These phases are implemented in the tight-binding model by multiplying all hopping terms in the positive $x$, $y$, and $z$ directions with phases $\lambda_x$, $\lambda_y$, and $\lambda_z$, respectively, where
\begin{subequations}\label{FluxesCHI}
	\begin{gather}
	\lambda_x = e^{i \frac{\Phi_x}{2N_x}},\\
	\lambda_y = e^{i \frac{\Phi_y}{2N_y}},\\
	\lambda_z = e^{i \frac{\Phi_z}{N_z}}.
	\end{gather}
\end{subequations}
Correspondingly, hopping terms with a component in the negative $x$, $y$, and $z$ directions are multiplied with the complex conjugate phases $\lambda_x^*$, $\lambda_y^*$, and $\lambda_z^*$. 

		\subsection{Hinge state characterization\label{sec:HingeStatesCHI}}
		
		In the CHI ground state with open boundary conditions in the $xy$-directions, each of the four hinges parallel to the $z$-axis supports a single chiral mode localized at the hinge~\cite{schindler2018higher}. Since the CHI model is non-interacting, each hinge mode is expected to correspond to one free bosonic mode described by a CFT with central charge $c=1$ and Luttinger parameter $K=1$, analogous to the edge states of a non-interacting Chern insulator with Chern number $\mathcal{C} = 1$. In order to confirm this expectation, we numerically extract $c$ and $K$ from the EE and particle number fluctuations using the same geometry as in the main text for the FCHI hinge modes, see Fig.~\ref{fig:SketchCHI}(b). 
		This geometry is a 3D generalization of the ``ribbon'' geometry used for similar analyses of 2D edge states~\cite{crepel2019model, crepel2019microscopic,estienne2019entanglement}, see Fig.~\ref{fig:EdgeModes}(a) and Appendix~\ref{sec:FCI_EdgeModes}. In both cases the cut is perpendicular to the hinge/edge modes, which results in contribution of these modes to the EE and the particle/spin fluctuations.
		
\begin{figure*}[t]
			\begin{tikzpicture}
			\node at (0, 0) {\includegraphics[width = 0.49\linewidth] {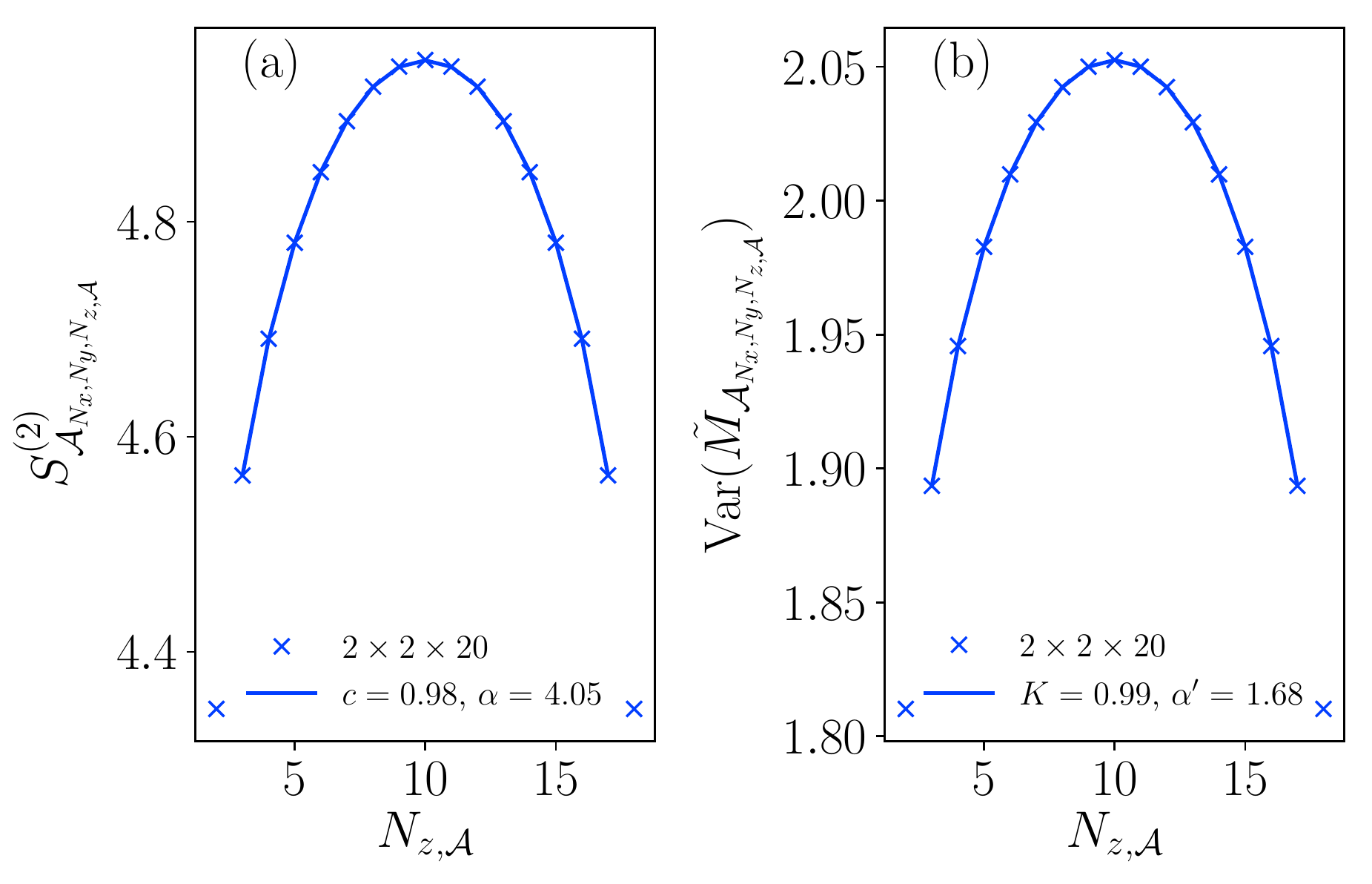}};
			\node at (9, 0) {\includegraphics[width = 0.49\linewidth] {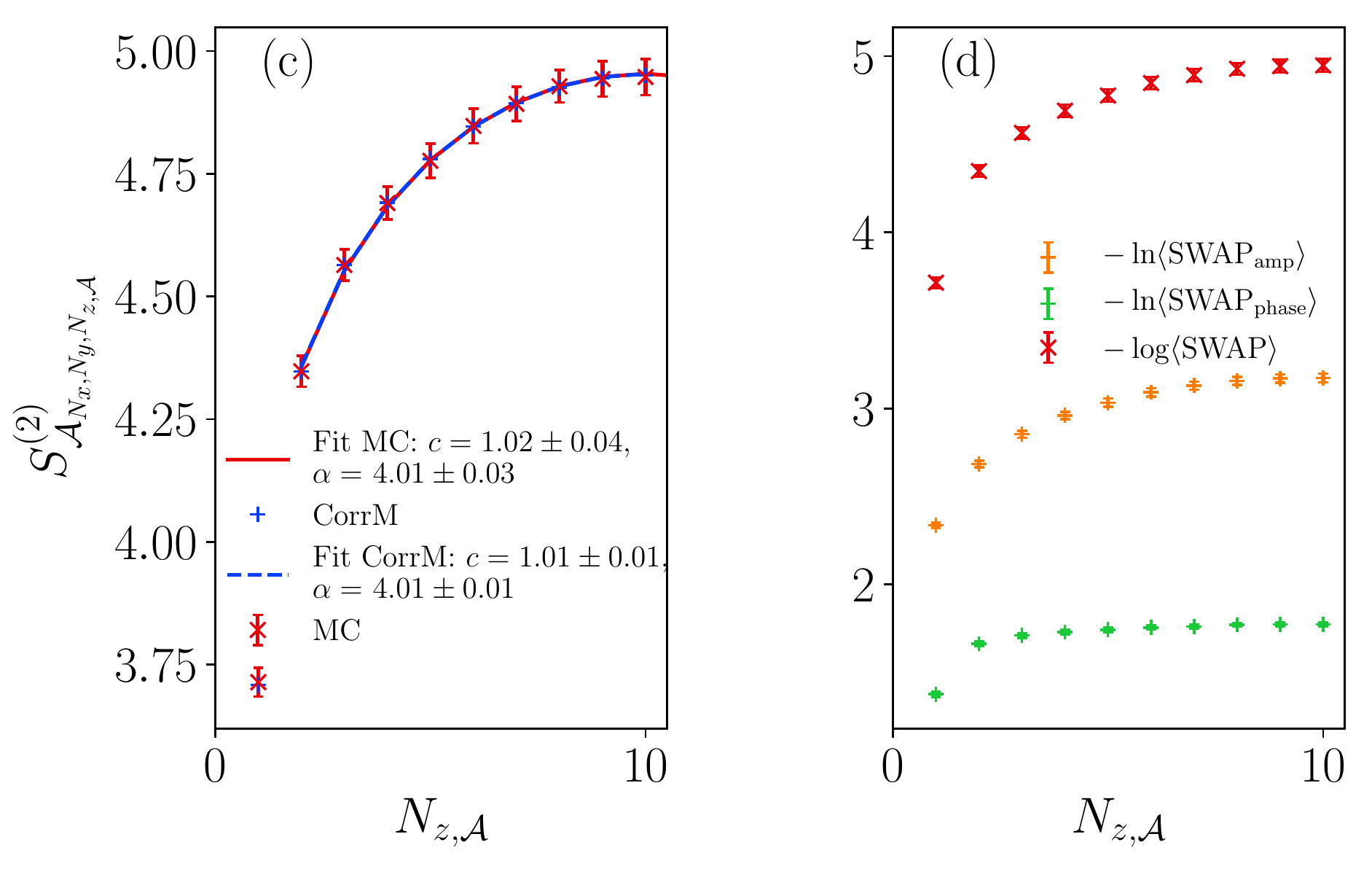}};
			\end{tikzpicture}	
			\caption[Hinge data CHI]{Numerical extraction of the central charge $c$ and Luttinger parameter $K$ of the CFT describing the chiral hinge modes of the non-interacting CHI. In (a), scaling of the second Renyi entropy as computed from the CorrM technique fit to the prediction of Eq.~\eqref{EEHingeAppC}, and in (b), scaling of the particle number fluctuations as computed from the CorrM technique fit to the prediction of Eq.~\eqref{FitFunctionParticleNumberVariance3DAppC} \wrt the series of subsystems $\mathcal{A}_{N_x, N_y, N_{z, \mathcal{A}}}$ as a function of $N_{z, \mathcal{A}}$. In (c), comparison of the MC results for the EE with those from the CorrM technique. In (d), formation of the MC result for the EE by addition of the contributions from the amplitude and phase of the wave function (\cf Appendix~\ref{sec:AppendixMC}). In all cases, the system has open boundary conditions and $N_x = 2$, $N_y = 2$ unit cells in the $x$, $y$ directions, and periodic boundary conditions and $N_z = 20$ unit cells in the $z$ direction.
			\label{fig:HingeDataCHI}}
		\end{figure*}

		Since the CHI is a free-fermion model, the EE and particle number fluctuations can be computed efficiently using the correlation matrix (CorrM) 
		method for free fermions. The results are shown in Fig.~\ref{fig:HingeDataCHI}(a) for the second Renyi entropy $S^{(2)}$, which is expected to obey the scaling of Eq.~\eqref{EEHinge} with central charge $c = 1$,
		\begin{equation}\label{EEHingeAppC}
		S^{(2)}_{\mathcal{A}_{N_x, N_y, N_{z, \mathcal{A}}}} (N_{z, \mathcal{A}}) = \alpha + 4 \times S^{(2)}_{\text{crit}} (N_{z, \mathcal{A}}; N_z), 
		\end{equation}
		where $S^{(2)}_{\text{crit}} (N_{z, \mathcal{A}}; N_z)$ is given by Eq.~\eqref{EE1DCritical} and the factor 4 comes from the number of hinge modes. 
		From Fig.~\ref{fig:HingeDataCHI}(a) we see that the data agree very well with this prediction, where the observed value for the central charge is $c = 0.98$. The numerical value for $c$ is in even closer agreement with the expected value $c = 1$ for bigger system sizes.
		
		Analogously to the main text and Appendix~\ref{sec:FCI_EdgeModes} where we considered the variance $\var(M_{\mathcal{A}})$~\eqref{eq:SpinFluctuations} of the number $M_{\mathcal{A}}$ of spin up particles in the region $\mathcal{A}$, we now consider the variance of the number of particles $\tilde{M}_{\mathcal{A}}$ (note that the particles are ``spinless'' in the non-interacting CHI model and we have a single copy):
		\begin{equation}
		\var(\tilde{M}_{\mathcal{A}_{N_x, N_y, N_{z, \mathcal{A}}}}) \equiv \langle \tilde{M}_{\mathcal{A}_{N_x, N_y, N_{z, \mathcal{A}}}}^2 \rangle - \langle \tilde{M}_{\mathcal{A}_{N_x, N_y, N_{z, \mathcal{A}}}} \rangle ^2.
		\end{equation}
		The Luttinger parameter $K$ for the hinge modes of the CHI can be extracted from the scaling of the particle number fluctuations.
		Note that for the non-interacting CHI, the particle number fluctuations $\var(\tilde{M}_{\mathcal{A}_{N_x, N_y, N_{z, \mathcal{A}}}})$ give access to a conserved current analogous to the spin fluctuations $\var(M_{\mathcal{A}_{N_x, N_y, N_{z, \mathcal{A}}}})$ for the fractional CHI as studied in the main text. Therefore, $\var(\tilde{M}_{\mathcal{A}_{N_x, N_y, N_{z, \mathcal{A}}}})$ in the CHI is expected to scale according to Eq.~\eqref{FitFunctionParticleNumberVariance3D} with Luttinger parameter $K = 1$:
		\begin{equation}\label{FitFunctionParticleNumberVariance3DAppC}
		\var(\tilde{M}_{\mathcal{A}_{N_x, N_y, N_{z, \mathcal{A}}}}) = 2 \times \frac{K} {\pi ^ 2}  \ln \left[ \frac{N_z}{\pi} \sin \left(\frac{\pi  N_{z, \mathcal{A}}} {N_z} \right)\right] + \alpha'.
		\end{equation}	
		From Fig.~\ref{fig:HingeDataCHI}(b) we see that the data agree very well with this expectation, where the observed value of the Luttinger parameter $K = 0.99$ is very close to the expected value $K = 1$.
		
		Moreover, we can use this geometry to benchmark our MC algorithm on a 3D system of similar size and nature as the FCHI studied in the main text. To that end, we compare the results for the EE obtained from the CorrM computation to those obtained from the MC algorithm, as sketched in Fig.~\ref{fig:HingeDataCHI}(c) for a system of size $2\times 2 \times 20$. It is clear that the results obtained from the two techniques are in very good agreement. Furthermore, in Fig.~\ref{fig:HingeDataCHI}(d) we show how the MC result for the entanglement entropy is obtained by addition of the contributions stemming from the amplitude and the phase of the wave function (\cf Appendix~\ref{sec:AppendixMC}). For the CHI, the amplitude contribution is dominant and contributes more significantly to the logarithmic hinge scaling. 
		
		Note that the computation of the EE via MC in this geometry is feasible only for systems which have a small cross section in the $xy$ plane. Indeed, for larger cross sections, the area law contribution contained in the constant $\alpha$ quickly grows such that the EE can no longer be computed in MC due to exponentially long convergence times (\cf Appendix~\ref{sec:AppendixMC}). The data presented in Fig.~\ref{fig:HingeDataCHI} show that the correlation length of the CHI model is sufficiently small that a cross section of $2\times 2$ is already big enough to extract the universal properties of the hinge states, despite a finite hinge state hybridization which is estimated around 0.57 in a system of size $2 \times 2 \times N_z$ (assuming an exponential localization of the hinge modes in the $x$ and $y$ directions).

\section{Topological degeneracy of the Fractional Chiral Hinge Insulator\label{sec:AppendixFCHI}}

In this appendix, we present our results for the topological degeneracy of the fractional chiral hinge insulator (FCHI). We begin in Appendix~\ref{sec:AppFCHI_General} with a description of our approach and a discussion of topological degeneracy for 3D systems. We then proceed to present our results for the topological degeneracy of the FCHI in isotropic geometries in Appendix~\ref{sec:AppFCHI_Isotropic}, and for anisotropic geometries with $N_z > N_x, N_y$ in Appendix~\ref{sec:AppFCHI_Anisotropic}. Finally, in Appendix~\ref{sec:AppFCHI_OBC} we discuss the topological degeneracy of the FCHI with open boundary conditions in the $x$ direction and periodic boundary conditions in the $y$ and $z$ directions. 

\subsection{Topological degeneracy for 3D systems\label{sec:AppFCHI_General}}

	In order to characterize the topological degeneracy of the FCHI, we consider a set of ansatz states obtained by Gutzwiller projection of the non-interacting CHI wave function with different boundary conditions for the underlying electronic degrees of freedom, and compute the number of linearly independent states among them. We are following the same procedure described in Appendix~\ref{sec:FCI_TopDeg} for a FCI. Concretely, to compute the topological degeneracy on the three-torus we consider the non-interacting wave function with periodic boundary conditions (PBC) or anti-periodic boundary conditions (APBC) in each direction. Indeed, both PBC and APBC for fermions lead to PBC for the Gutzwiller projected state. This yields eight ansatz states for the FCHI on the three-torus. 
	
	In keeping with the notation from the main text, we denote by $\ket{\psi^{(\Phi_x, \Phi_y, \Phi_z)}_{s}}$ the ground state of the non-interacting CHI model with twisted boundary conditions $\Phi_x , \Phi_y, \Phi_z  \in \{0, \pi\}$ corresponding to PBC and APBC, respectively, and with spin $s \in \{\uparrow, \downarrow\}$. The fractional wave function obtained by Gutzwiller projection of two copies of the non-interacting wave function with the same flux insertions but with opposite spin is denoted
	\begin{equation}
	\ket{\Psi^{(\Phi_x, \Phi_y, \Phi_z)}} = P_G\left[\ket{\psi^{(\Phi_x, \Phi_y, \Phi_z)}_{\uparrow}} \otimes \ket{\psi^{(\Phi_x, \Phi_y, \Phi_z )}_{\downarrow}}\right].
	\end{equation}
	In order to determine the topological degeneracy, we need to compute the rank of the overlap matrix $\mathcal{O}$, whose entries
	\begin{equation}
	\mathcal{O}_{(\Phi_x, \Phi_y, \Phi_z), (\Phi_x', \Phi_y',\Phi_z')} = \frac{\myexp{	\Psi^{(\Phi_x, \Phi_y, \Phi_z)} | \Psi^{(\Phi'_x, \Phi'_y, \Phi'_z)} }} {\sqrt{\myexp{	\Psi^{(\Phi_x, \Phi_y, \Phi_z)} | \Psi^{(\Phi_x, \Phi_y, \Phi_z)} } } \sqrt{\myexp{	\Psi^{(\Phi'_x, \Phi'_y, \Phi'_z)} | \Psi^{(\Phi'_x, \Phi'_y, \Phi'_z)} } }}.
	\end{equation}
	contain the overlap between the normalised ansatz states for the FCHI. Since we start with eight ansatz states, the rank of the overlap matrix can be at most equal to eight. In the following, we study if this maximal rank is saturated or if there are linear dependencies between the ansatz states leading to a reduction of its rank.

\subsection{Isotropic case\label{sec:AppFCHI_Isotropic}}

\subsubsection{Stability under staggered chemical potential $\mu$ for $2\times 2 \times 2$}

\begin{figure*}[t]
	\begin{tikzpicture}
	\node at (0, 0) {\includegraphics[width = 0.5\linewidth] {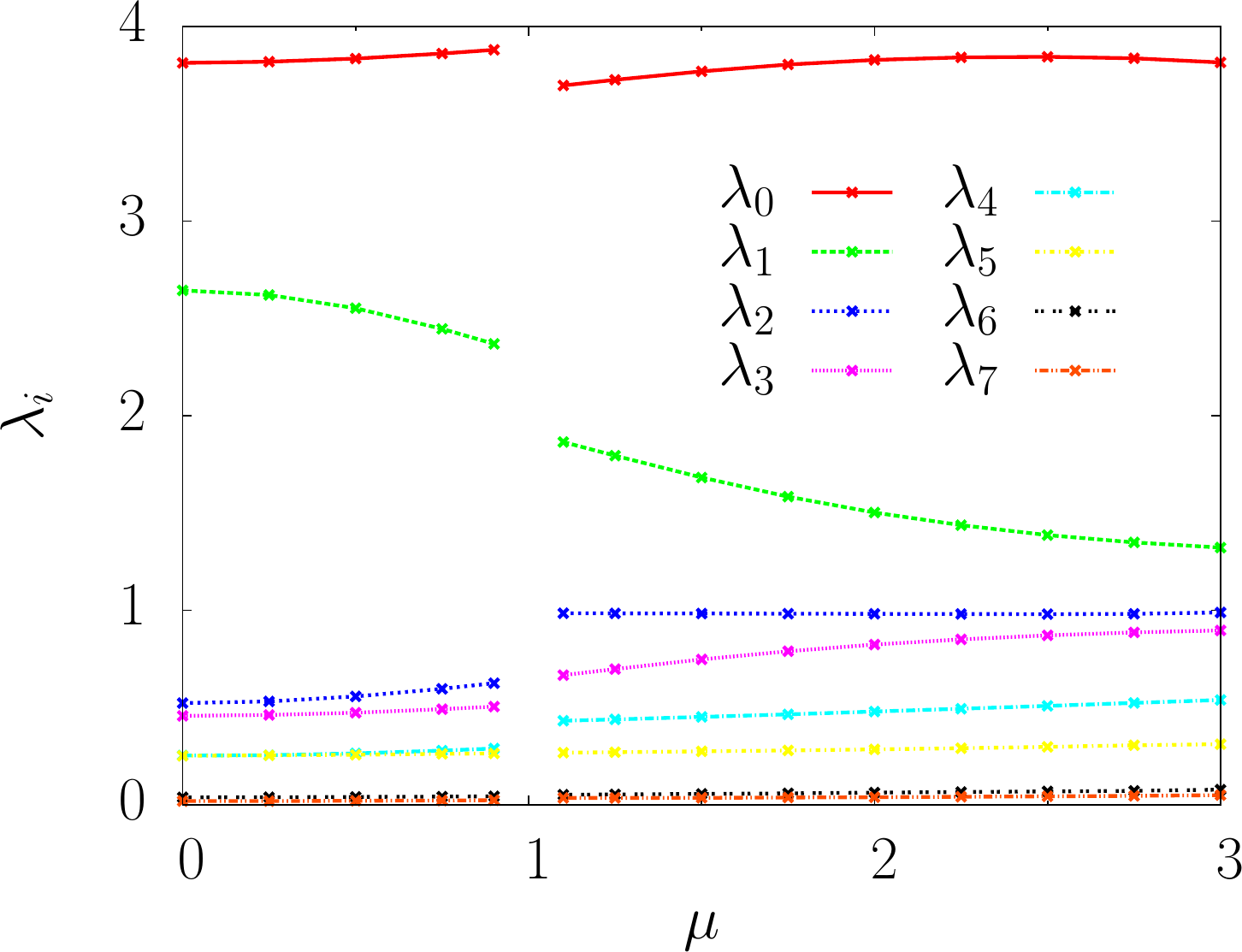}};
	\end{tikzpicture}	
	\caption[FCHI Overlap matrix eigenvalues]{
		Overlap matrix eigenvalues ($\lambda_0$ to $\lambda_7$) as a function of staggered chemical potential $\mu$ for the FCHI on a three-torus. The system size is $2\times 2\times 2$ unit cells.
		\label{fig:FCHI_OverlapEV_SameMu}}
\end{figure*}

As we already did for the FCI in Appendix~\ref{sec:FCI_TD_mu}, we now consider the effect of a staggered chemical potential $\mu$ as defined in Eq.~\eqref{BlochHamiltonianChiralHingeTIStaggeredMu} on the eigenvalues of the overlap matrix for the FCHI on a three-torus. Here, the chemical potential $\mu$ is the same in both copies of the non-interacting CHI underlying the FCHI wave function. We study a system of size $2\times 2 \times 2$ unit cells and perform ED. The computation of the overlap matrix elements is simpler using ED compared to MC, as previously discussed in Appendix~\ref{FCI_TD_mu0}.

The overlap matrix eigenvalues as a function of the staggered chemical potential $\mu$ are shown in Fig.~\ref{fig:FCHI_OverlapEV_SameMu}. Similar to the case of the FCI, most of the eigenvalues are obviously discontinuous at $\mu_c=1$. This is the critical point where the underlying CHI model becomes trivial, see Appendix~\ref{sec:HamiltonianCHI}. This is a very small system and there are no exact degeneracies, therefore it is expected that the finite size overlap matrix rank is equal to the maximal value 8. However, the separation between sets of eigenvalues is not as clear as for the FCI case, albeit the two largest eigenvalues stand out from the rest. For the FCHI case we cannot explore the scaling of overlap matrix eigenvalues with the system size using ED, as $2\times 2\times 2$ is the largest system accessible to this method. Larger systems will be studied in the following sections using MC computations.

We again emphasize that the Gutzwiller projection in the limit of large $\mu$ might not be meaningful: all the particles are (mostly) located on the same sites (3 and 4) in both CHI copies and the Gutzwiller projection excludes double occupancies thus leaving only particle fluctuations. This was discussed in more details in Appendix~\ref{sec:FCI_TD_mu}. In particular, we have not investigated the opposite staggered chemical potential for FCHI as we have shown this was not physically meaningful in the ``topological'' regime for the FCI (see Appendix~\ref{sec:FCI_TEE}).

\subsubsection{Larger systems at $\mu=0$}

\begin{figure}[t]
	\includegraphics[width = 0.5\linewidth] {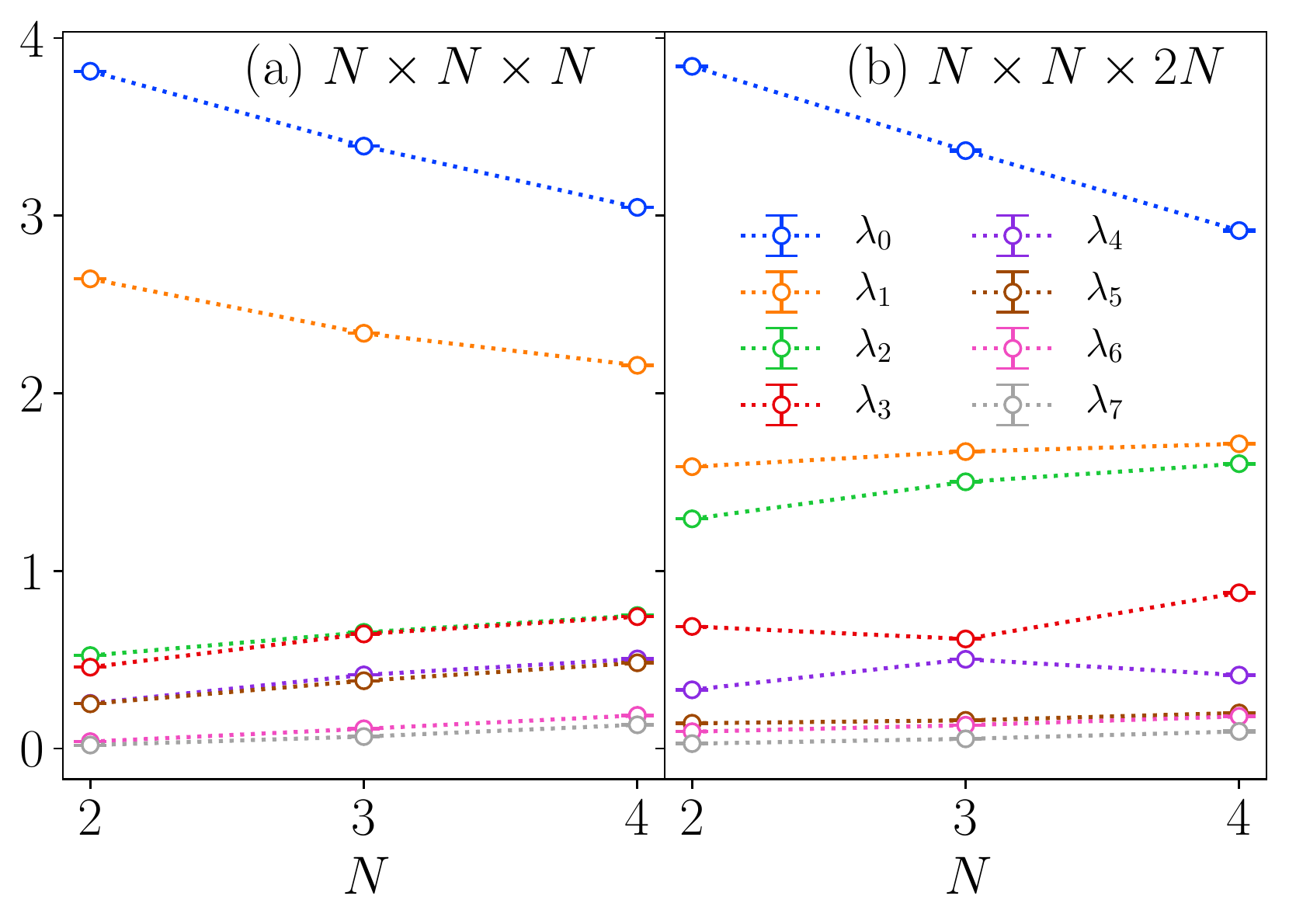}
	\caption[Overlap matrix eigenvalues FCHI isotropic scaling]{Scaling of the overlap matrix eigenvalues for the FCHI on a three-torus for isotropic systems of size $N \times N \times N$ in (a) and size $N \times N \times 2N$ in (b).\label{fig:OverlapFCHIIsotropic}}
\end{figure}

We now move on to bigger systems, which are accessible only via MC computations. Focusing solely on $\mu=0$, we attempt to study the overlap matrix eigenvalues of the FCHI if we increase the system size in an isotropic fashion. We consider two cases: on one hand, a system of $N \times N \times N$ unit cells, and on the other hand, a system of $N \times N \times 2N$ unit cells (where the number of lattice sites in each direction is equal since the unit cell contains two sites in the horizontal $x$ and $y$ directions, but only a single site in the $z$ direction). The results for the eigenvalues of the overlap matrix as computed from MC are sketched in Fig.~\ref{fig:OverlapFCHIIsotropic}(a) and (b) for these two cases, respectively. Due to the 3D setting and the large number of observables that we have to compute to obtain the full overlap matrix (\cf Appendix~\ref{sec:AppendixMC}), we are restricted to relatively small systems up to $N =4$. For the biggest of these systems, the computations already consumed a considerable number of CPU hours (see Appendix~\ref{sec:AppendicMCTechnical}).
 
From the data presented in Fig.~\ref{fig:OverlapFCHIIsotropic} we cannot infer a non-trivial reduction of the number of linearly independent states for the FCHI in the thermodynamic limit for isotropic systems. Indeed, it appears that the largest eigenvalue is decreasing as $N$ increases, whereas the smaller eigenvalues appear to increase. This would indicate that the number of linearly independent states is equal to 8 which is the maximal number given the size of the overlap matrix. However, it is possible that this is a finite size effect and that the result differs for bigger systems which are not accessible in numerical computations.

\subsection{Anisotropic case\label{sec:AppFCHI_Anisotropic}}

\begin{figure}[t]
	\includegraphics[width = 0.99\linewidth] {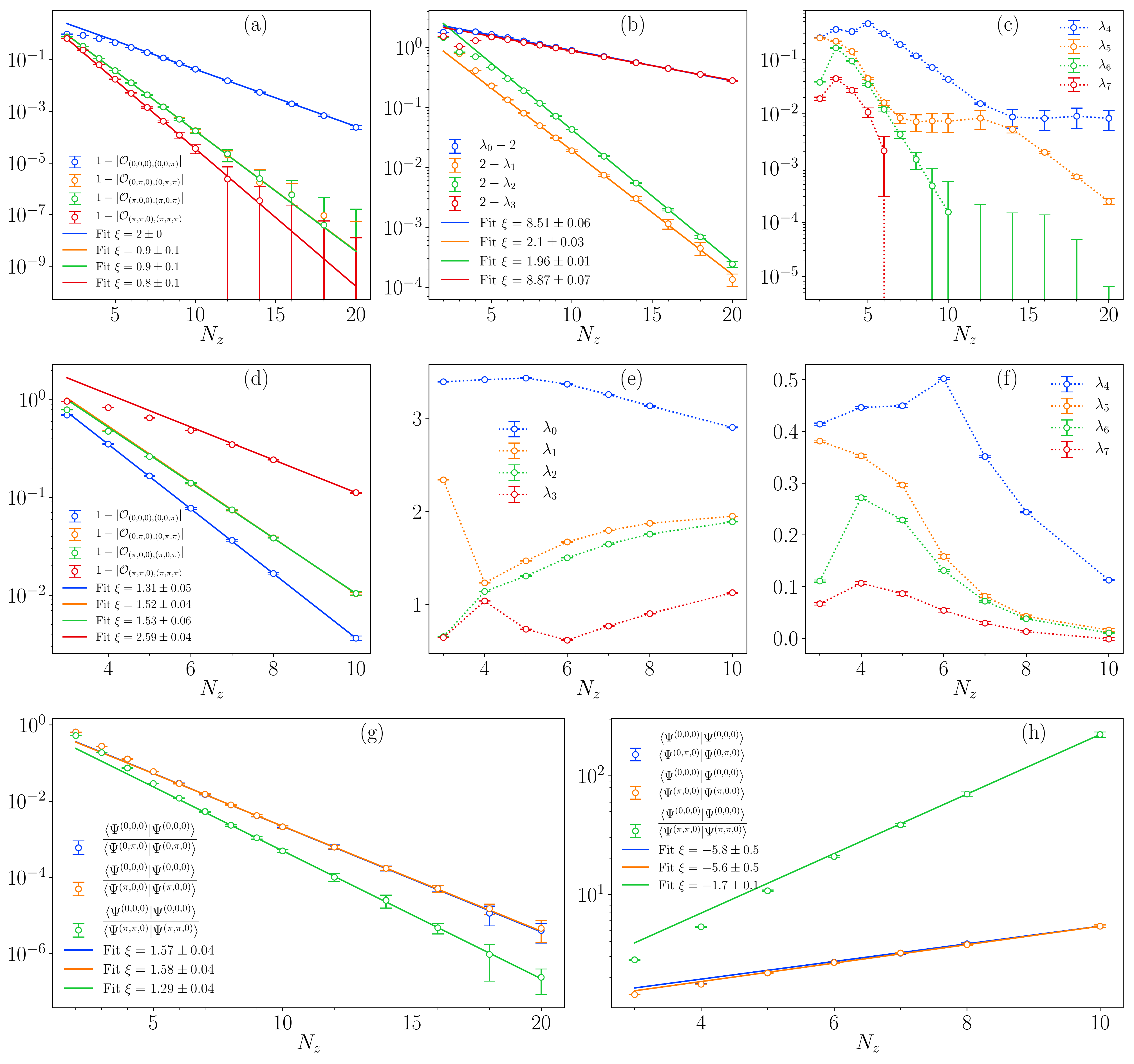}	
	\caption[Overlap matrix FCHI anisotropic scaling]{Characterization of the overlap matrix in the anisotropic limit for systems of size $2 \times 2 \times N_z$ in (a), (b), (c) and (g), and for systems of size $3\times 3 \times N_z$ in (d), (e), (f) and (h). Where possible, we have fit the data to an exponential decay of the form $u e^{-i N_z / \xi}$ with a ``correlation length'' $\xi$ in units of unit cells and an amplitude $u$ (not shown). In (a) and (d), we find an exponentially fast approach to the value 1 of the normalised overlap between states differing only by a $\pi$-flux in the $z$ direction, showing that the states become identical in the limit $N_z \rightarrow \infty$. In (b) and (e), scaling of the four largest eigenvalues of the overlap matrix as a function of $N_z$. Note that in (b), we show the difference from the value 2 of the eigenvalues on a logarithmic scale to demonstrate the exponential approach to the value 2. In (c) and (f), scaling of the four smallest eigenvalues of the overlap matrix. In (g) and (h), ratio of the norms of the ansatz states with vanishing flux in the $z$ direction after Gutzwiller projection but before normalization. For $N_x, N_y$ even, the norms of the states with zero or only a single $\pi$-flux in the horizontal directions are exponentially suppressed compared to the weight of the states with $\pi$-fluxes in both horizontal directions. For $N_x, N_y$ odd, the weights of states with at least one $\pi$-flux in the horizontal directions are  exponentially suppressed compared to the weight of the state without any $\pi$-fluxes. Note that the exponential growth in (h) corresponds to negative values of the fitted correlation length $\xi$.\label{fig:OverlapFCHIAnisotropic}}
\end{figure}

In addition to the isotropic case, we also studied the topological degeneracy of the FCHI for anisotropic systems,  where $N_x = N_y$  such that the $C_4$ rotation symmetry in the horizontal plane is preserved, but where $N_z$ is larger than $N_x$ and $N_y$. Note that this is the aspect ratio which we used in the main text to study the hinge mode physics, albeit  with open instead of periodic boundary conditions in the horizontal directions. 

In contrast to the isotropic case, for anisotropic systems the ground state degeneracy is reduced if $N_z$ is much larger than $N_x$ and $N_y$. Indeed, the normalized overlap between the two ansatz states $\ket{\Psi^{(\Phi_x, \Phi_y, 0)}}$ and $\ket{\Psi^{(\Phi_x, \Phi_y, \pi)}}$, with the same fluxes $(\Phi_x, \Phi_y)$ in the horizontal directions but differing flux in the $z$ direction, approaches unity \emph{exponentially} fast as $N_z$ increases. This is shown in Fig.~\ref{fig:OverlapFCHIAnisotropic}(a) and (d) for systems of size $2 \times 2 \times N_z$ and $3\times 3 \times N_z$, respectively. This implies that the two ansatz states $\ket{\Psi^{(\Phi_x, \Phi_y, 0)}}$ and $\ket{\Psi^{(\Phi_x, \Phi_y, \pi)}}$ become linearly dependent in the limit $N_z \rightarrow \infty$. Since there are four different flux patterns $(\Phi_x, \Phi_y)$ in the horizontal directions, in the limit $N_z \rightarrow \infty$ the 8 ansatz states split into four pairs, where the two states in each pair have the same $(\Phi_x, \Phi_y)$ and are linearly dependent. 

\begin{figure}[t]
	\includegraphics[width = 0.5\linewidth] {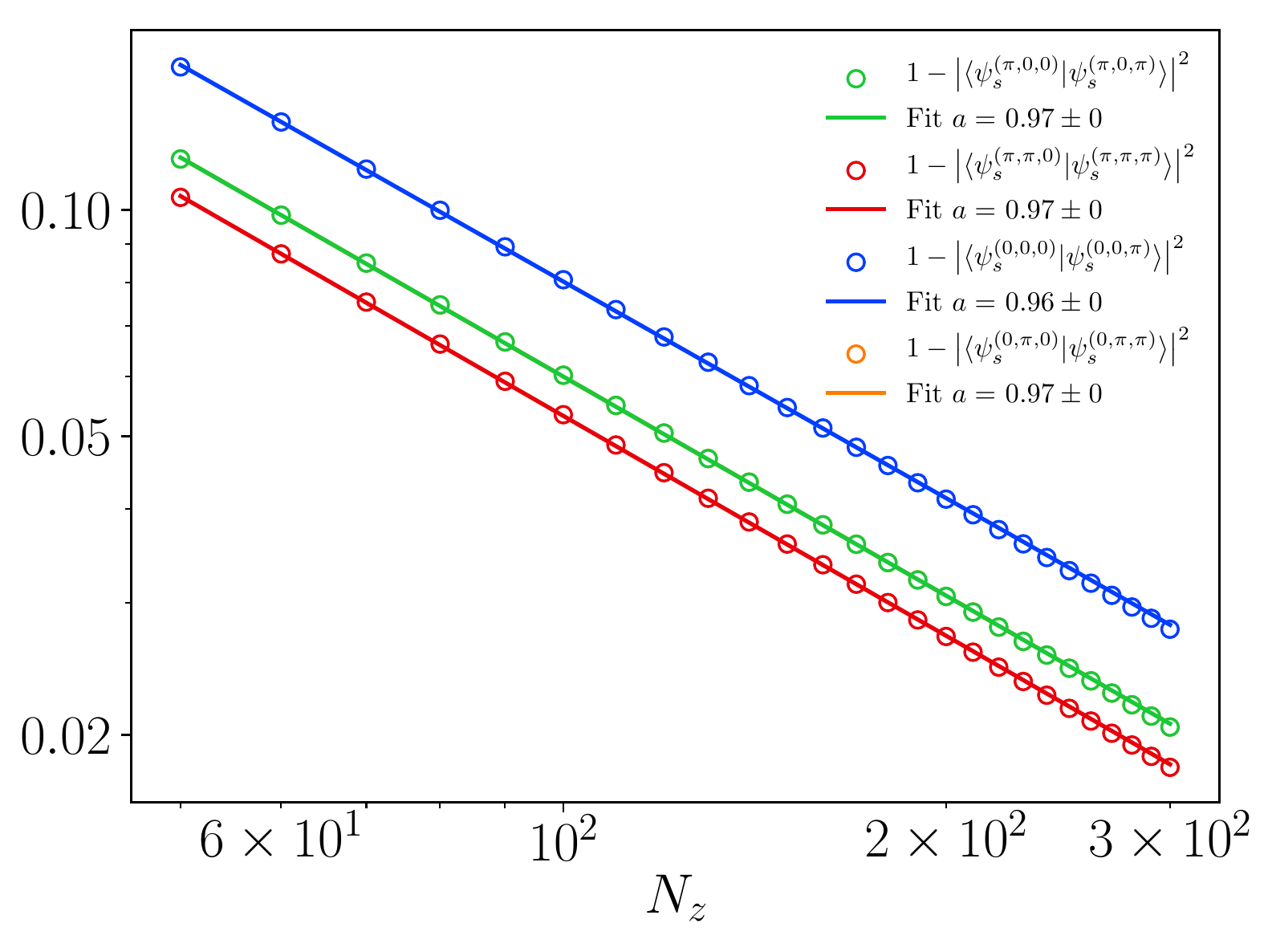}	
	\caption[Overlaps CHI anisotropic scaling]{Difference from unity of the squared overlap $|\langle\psi^{(\Phi_x, \Phi_y, 0)}_{s} | \psi^{(\Phi_x, \Phi_y, \pi)}_{s}\rangle|^2$ between the normalised many-body non-interacting CHI ground states with the same fluxes $(\Phi_x, \Phi_y)$ in the horizontal directions but differing flux in the $z$ direction in the anisotropic limit for a system of size $2 \times 2 \times N_z$. The data is fit to an algebraic decay $b \times N_z^{-a}$ with power $a$. \label{fig:OverlapsAnisotropicCHI}}
\end{figure}

We note that a similar phenomenon occurs for the non-interacting CHI whose ground states generate the FCHI ansatz states by Gutzwiller projection. Indeed, we have checked for several fixed values of $N_x = N_y$ that the overlap
\begin{equation}
\left|\myexp{\psi^{(\Phi_x, \Phi_y, 0)}_{s} | \psi^{(\Phi_x, \Phi_y, \pi)}_{s}}\right|^2
\end{equation}
between the normalised many-body non-interacting ground states with the same fluxes $(\Phi_x, \Phi_y)$ in the horizontal directions but differing flux in the $z$ direction also approaches unity as $N_z$ increases. One example for systems of size $2 \times 2 \times N_z$ is shown in Fig.~\ref{fig:OverlapsAnisotropicCHI}. (Note that for two non-interacting many-body states $\ket{\psi^{(\Phi_x, \Phi_y, 0)}_{s}}$ and $\ket{\psi^{(\Phi_x, \Phi_y, \pi)}_{s}}$ their square overlap can easily be computed as the determinant of the sum of their correlation matrices). This implies that the two states $\ket{\psi^{(\Phi_x, \Phi_y, 0)}_{s}}$ and $\ket{\psi^{(\Phi_x, \Phi_y, \pi)}_{s}}$ become identical in the limit $N_z \rightarrow\infty$. However, we have observed that for the non-interacting model this convergence to the value one is \emph{algebraic} and thus much slower than for the FCHI, where the overlap approaches one exponentially. Therefore, we believe that the behavior shown in Fig.~\ref{fig:OverlapFCHIAnisotropic}(a) and (d) is qualitatively new and reserved to the interacting wave function.

As a result of the linear dependencies between the FCHI ansatz states for $N_z$ much larger than $N_x$ and $N_y$, the ground state degeneracy of the FCHI in this case can be at most four. As shown in Fig.~\ref{fig:OverlapFCHIAnisotropic}(b) and (e), the four larger eigenvalues of the overlap matrix approach a non-zero value as $N_z$ increases, implying that the four different horizontal flux combinations generate four linearly independent ansatz states. On the other hand, the four smaller eigenvalues go to zero exponentially fast with increasing $N_z$ as shown in Fig.~\ref{fig:OverlapFCHIAnisotropic}(c) and (f). It is interesting to note that the four largest eigenvalues all approach the same value $\lambda = 2$ exponentially. This is a similar behavior as for the FCI, where the two non-zero eigenvalues also approach the same value in the thermodynamic limit (\cf Appendix~\ref{sec:FCI_TopDeg}). However, we have not found any arguments indicating that the asymptotic degeneracy of all non-zero eigenvalues of the overlap matrix contains information on the topology of a system.

In the discussion above, we have always considered the normalized overlap matrix, which measures the topological degeneracy on the manifold of \emph{normalized} ansatz states. In other words, we have defined different ansatz states for the interacting model by projecting the wave function of the non-interacting model, normalizing each ansatz state separately and only then considering linear dependencies. 

However, it may also be valid to follow a different approach where one considers the linear independence of the \emph{unnormalized} ansatz states for the interacting model. In other words, one considers linear combinations of the ansatz states after the Gutzwiller projection but before normalization. In the thermodynamic limit, this may lead to a different result for the rank of the overlap matrix if the different unnormalized ansatz states for the interacting model have very different weights. Indeed, this is the case here as shown in Fig.~\ref{fig:OverlapFCHIAnisotropic}(g) and (h) for systems of size $2 \times 2 \times N_z$ and $3\times 3 \times N_z$, respectively. In both cases, there is one out of the four horizontal flux combinations, denoted $(\Phi_x^0, \Phi_y^0)$, for which the corresponding ansatz states $\ket{\Psi^{(\Phi_x^0, \Phi_y^0, \Phi_z)}}$ have a weight which grows exponentially with increasing $N_z$ compared to the weights of the ansatz states obtained for the other three horizontal flux combinations. Note that the weights of $\ket{\Psi^{(\Phi_x^0, \Phi_y^0, 0)}}$ and $\ket{\Psi^{(\Phi_x^0, \Phi_y^0, \pi)}}$ are asymptotically identical. This dominant horizontal flux combination $(\Phi_x^0, \Phi_y^0)$ is staggered in $N_x$ and $N_y$, where
\begin{equation}
(\Phi_x^0, \Phi_y^0) = \begin{cases}
(\pi, \pi) \qquad &\text{for $N_x = N_y$ even}\\
(0, 0) \qquad &\text{for $N_x = N_y$ odd}
\end{cases}.
\end{equation}

Following this approach, the ground state degeneracy is given by the rank of a rescaled overlap matrix $\tilde{\mathcal{O}}$ with entries
\begin{equation}
\tilde{\mathcal{O}}_{(\Phi_x, \Phi_y, \Phi_z), (\Phi_x', \Phi_y',\Phi_z')} = \frac{\myexp{	\Psi^{(\Phi_x, \Phi_y, \Phi_z)} | \Psi^{(\Phi'_x, \Phi'_y, \Phi'_z)} }} {\myexp{	\Psi^{(\Phi^0_x, \Phi^0_y, 0)} | \Psi^{(\Phi^0_x, \Phi^0_y, 0)} } }.
\end{equation}
Note that the trace of the overlap matrix of the unnormalized ansatz states is not normalized to eight. Here, we have therefore normalized $\tilde{\mathcal{O}}$ \wrt the weight of the state with dominant horizontal flux combination, which allows for a meaningful comparison between the eigenvalues of $\tilde{\mathcal{O}}$ for different system sizes. For the anisotropic FCHI, the rescaled overlap matrix $\tilde{\mathcal{O}}$ has one dominant eigenvalue converging to the value $\lambda = 2$, and seven eigenvalues decaying exponentially to zero with different correlation lengths. Therefore, following this approach, the FCHI has only one ground state for $N_z$ much larger than $N_x$ and $N_y$. Note that for the isotropic case discussed above, the spectrum of the rescaled overlap matrix $\tilde{\mathcal{O}}$ is very similar to that of $\mathcal{O}$ for the system sizes we have studied.

The large difference in the weight of the ansatz states that we observe for the anisotropic FCHI does not appear to be a necessary consequence of the reduction of the rank of the overlap matrix to a value lower than $8$ in the thermodynamic limit. For instance, for the FCI on the two-torus with $6 \times 6$ and $8 \times 8$ unit cells the construction discussed in Appendix~\ref{sec:FCI_TopDeg} yields four ansatz states with approximately the same weight, even though the rank of the overlap matrix is reduced from four to two. Therefore, for the FCI the normalised and the rescaled overlap matrix give the same result for the topological degeneracy.

	\begin{figure}[t]
	
	\includegraphics[width = \linewidth] {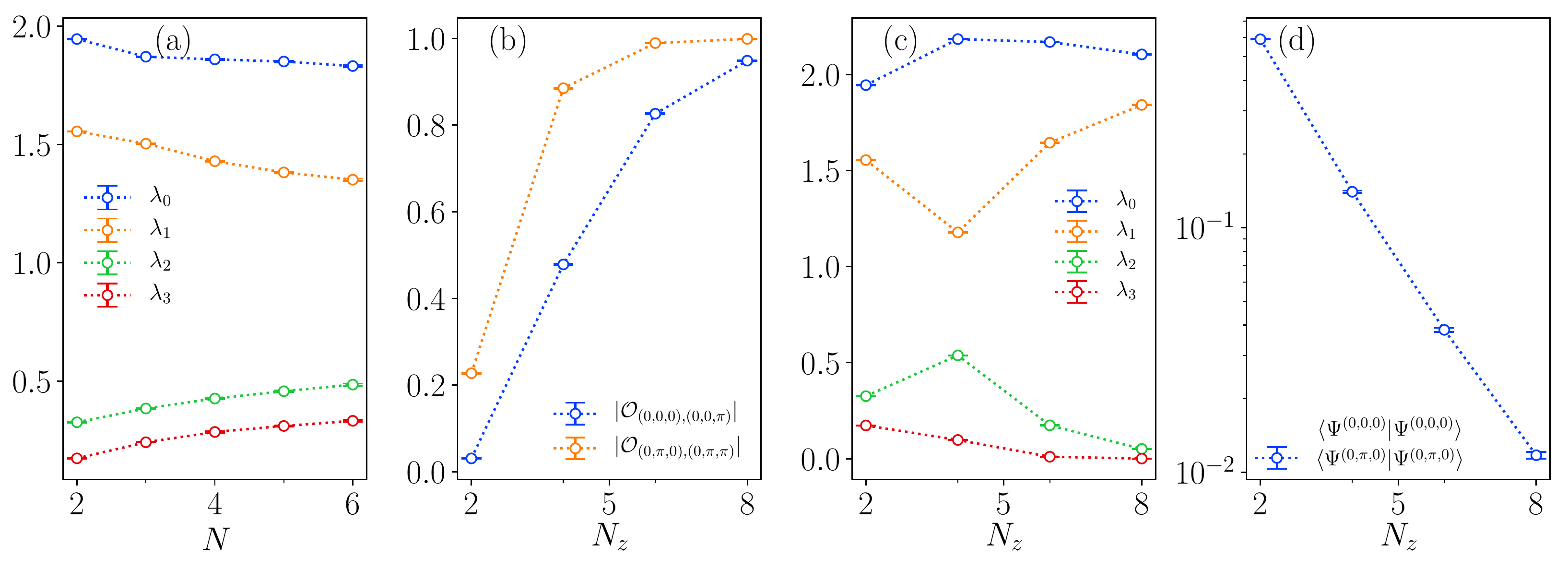}
	
	\caption[Overlap matrix OBC $\times$ PBC $\times$ PBC]{Ground state degeneracy of the FCHI with OBC in the $x$ direction and PBC in the $y$ and $z$ directions. In (a), overlap matrix eigenvalues for isotopic systems with $N \times N \times N$ unit cells. In (b) to (d), scaling of the overlap matrix for anisotropic systems of size $2 \times 2 \times N_z$ as a function of $N_z$. In (b), normalized overlap of states differing only by a $\pi$-flux in the $z$ direction approaching the value one in the limit $N_z \rightarrow\infty$. In (c), overlap matrix eigenvalues. In (d), ratio of the norms of the two states with different fluxes in the the $y$ direction and no flux in the $z$ direction.\label{fig:OverlapFCHIMisc}}
\end{figure}

\subsection{OBC in $x$\label{sec:AppFCHI_OBC}}

Finally, let us discuss the topological degeneracy of the FCHI with OBC in the $x$ direction and PBC in the other two directions. This is the configuration which we used in the main text to extract the TEE contributed by the gapped surface states. For these boundary conditions, we consider four ansatz states for the FCHI obtained by Gutzwiller projection of the non-interacting wave function with either PBC and APBC in the $y$ and $z$ directions. In the $x$ direction, all four ansatz states have vanishing flux. Therefore, the overlap matrix $\mathcal{O}$ now has dimension four.

The results for the topological degeneracy of the FCHI with this boundary configuration are very similar to those discussed above for the 3D torus. We first increase the system size in an isotropic fashion. 
While there is a separation between two larger and two smaller eigenvalues, they seem to converge to a finite value as shown in Fig.~\ref{fig:OverlapFCHIMisc}(a). This indicates that for isotropic systems the maximum rank of the overlap matrix is saturated. However, as before we are restricted to relatively small systems (albeit bigger that the all PBC case discussed in Appendix~\ref{sec:AppFCHI_Isotropic} due to the smaller number of overlaps to compute; note that the MC algorithm does not benefit from an intrinsic speedup due to PBC like an exact diagonalization would). Therefore we cannot make any reliable statements about the thermodynamic limit.

On the other hand, for anisotropic systems the normalized overlap between two ansatz states $\ket{\Psi^{(0, \Phi_y, 0)}}$ and $\ket{\Psi^{(0, \Phi_y, \pi)}}$, with the same flux $\Phi_y$ in the $y$ direction but differing flux in the $z$ direction, approaches unity as $N_z$ increases. This is shown in Fig.~\ref{fig:OverlapFCHIMisc}(b) for systems of size $2 \times 2 \times N_z$. Correspondingly, as shown in Fig.~\ref{fig:OverlapFCHIMisc}(c) the normalised overlap matrix for $N_z$ much larger than $N_x$ and $N_y$ has two finite eigenvalues converging to the value $\lambda = 2$, and two eigenvalues that vanish as $N_z$ increases. Again, the weight of the ansatz states before normalisation is not the same, with the weight of the states with $\Phi_y = 0$ being exponentially suppressed compared to the weight of the states with $\Phi_y = \pi$ for $N_x = N_y = 2$ as shown in Fig.~\ref{fig:OverlapFCHIMisc}(d). This implies that the overlap matrix $\tilde{\mathcal{O}}$ computed from the ansatz states before normalization has one finite and 3 vanishing eigenvalues for $N_z$ much larger than $N_x$ and $N_y$.

\afterpage{\clearpage}

\section{Technical data for MC computations\label{sec:AppendicMCTechnical}}
In this appendix, we provide some technical details on our MC simulations. In Appendix~\ref{sec:AppMCUpdate}, we discuss the update used in the simulations, and in Appendix~\ref{sec:AppendixRunTimes} we give the technical parameters and run times for all computations whose results are presented in the main text.

\subsection{Monte Carlo update\label{sec:AppMCUpdate}}

As discussed in the main text, charge fluctuations in the FCHI wave function are frozen out and the layer index (which will from now on be dubbed spin) is the only relevant degree of freedom on each site. The same holds for the FCI wave function studied in Appendix~\ref{sec:AppendixFCI}. Therefore, our MC computations are performed in the basis of spin configurations $\ket{v} = \ket{s_0, \dotsc, s_{N-1}}$ with $s_i \in \{\uparrow, \downarrow\}$ on each site $i = 0, \dotsc, N-1$, where $N$ is the total number of physical lattice sites. We used a single-spin-exchange update to suggest a new many-body configuration after each MC step. In other words, after each MC step, the configuration $\ket{v}$ is updated by exchanging the spin values $s_i$ and $s_j$ on two randomly chosen sites with opposite spin occupations $s_i \neq s_j$. 

In order to improve the acceptance rate of the simulations, we limited the range of the spin exchange to an integer value $r_{\text{update}}$. Concretely, we require that the graph distance $d(i,j)$ on the relevant lattice of the two sites $i, j$ should satisfy $d(i, j) \leq r_{\text{update}}$. Here, the relevant lattices are the cubic lattice for the FCHI and the square lattice for the FCI. Therefore, for $r_{\text{update}} = 1$ this permits spin exchange only between nearest-neighbour sites, whereas for $r_{\text{update}} = 2$ spin exchange both between nearest-neighbour and next-nearest-neighbour sites is allowed.

\begin{figure}
	\includegraphics[width = 0.5\linewidth] {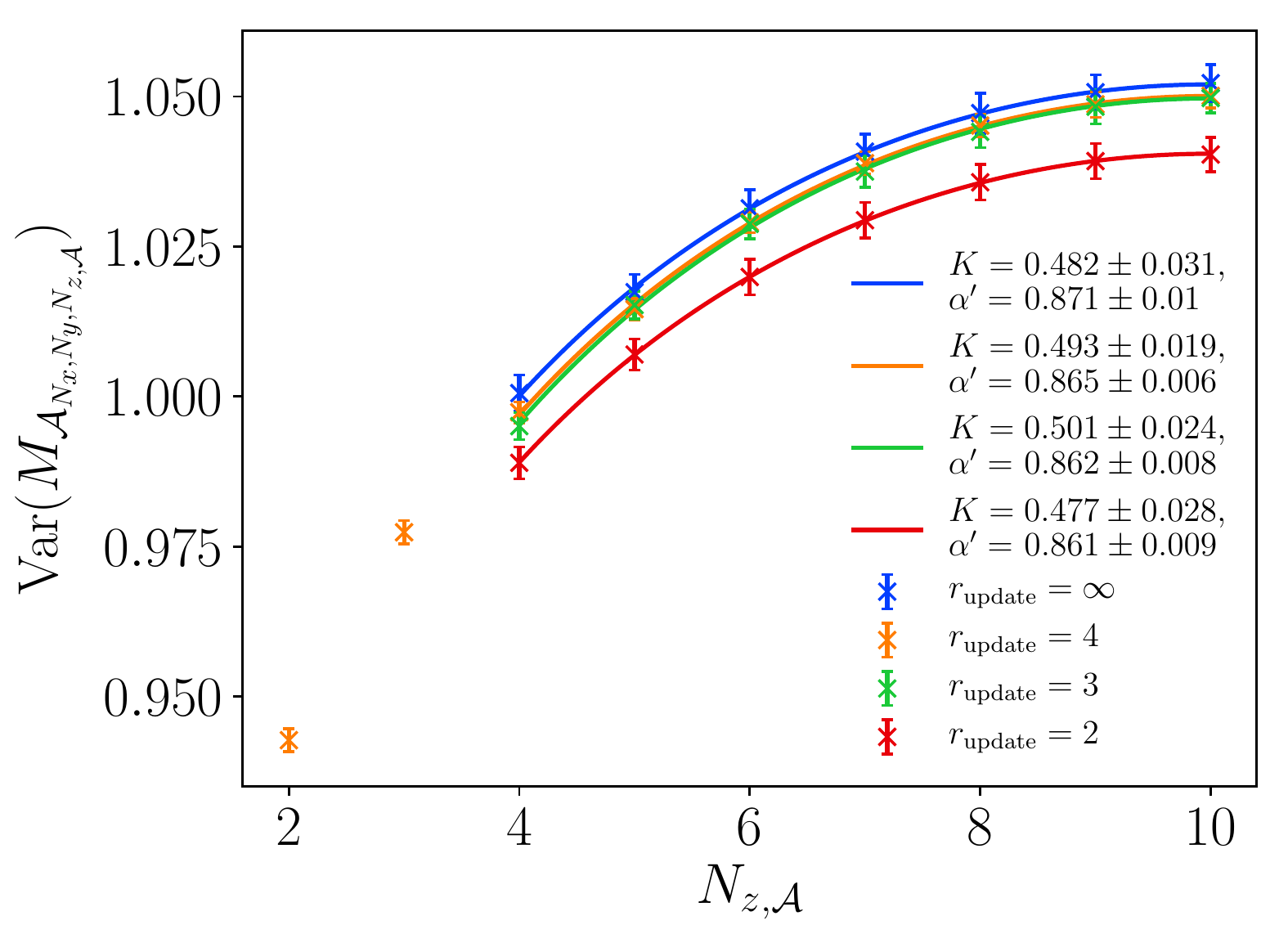}
	
	\caption[Comparison MC results different $r_{\text{update}}$]{MC results for computations with different $r_{\text{update}}$ for the spin number variance $\var (M_{\mathcal{A}_{N_x, N_y, N_{z, \mathcal{A}}}})$ in the subsystem $\mathcal{A}_{N_x, N_y, N_{z, \mathcal{A}}}$ of the FCHI as a function of $N_{z, \mathcal{A}}$, fit to the prediction of Eq.~\eqref{FitFunctionParticleNumberVariance3D}.\label{fig:ComparisonHoppingRange}}
\end{figure}

We have observed that the parameter $r_{\text{update}}$ has a small systematic influence on the mean value of the MC simulations. For instance, Fig.~\ref{fig:ComparisonHoppingRange} provides a comparison of the MC results for the spin fluctuations in the FCHI, as discussed in the main text, for different values $r_{\text{update}} = 2, 3, 4, \infty$. The data points for $r_{\text{update}} = 2$ do not agree within the statistical error bar with the data points for the most accurate measurement with $r_{\text{update}} = \infty$. However, for larger values $r_{\text{update}} = 3$ and $4$, the data points quickly move much closer to those for $r_{\text{update}} = \infty$ and their statistical error bars overlap. Indeed, the fit values for the parameters $K$ and $\alpha'$ agree within the statistical error bars for all four simulations. For all MC simulations presented in this paper, we chose $r_{\text{update}}$ sufficiently big that the systematic deviation is insignificant compared to the statistical uncertainty, while increasing the acceptance rate as much as possible.

\subsection{MC errors}

\begin{figure}
	\includegraphics[width = \linewidth] {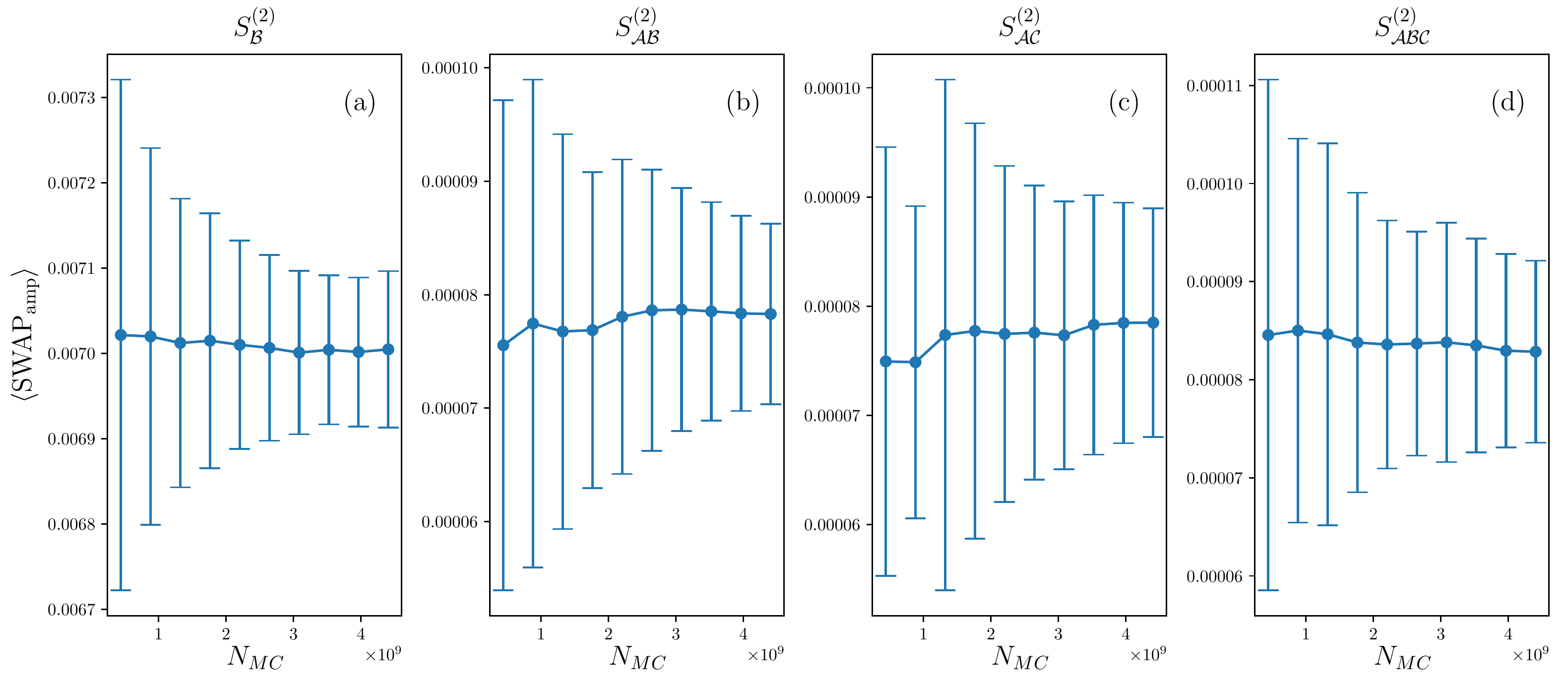}
	
	\caption[Convergence MC measurement]{Evolution of the mean value of the $\swapAmp$ observable for the four EE measurements $S^{(2)}_{\mathcal{B}}$ in (a), $S^{(2)}_{\mathcal{A}\mathcal{B}}$ in (b), $S^{(2)}_{\mathcal{A}\mathcal{C}}$ in (c) and $S^{(2)}_{\mathcal{A}\mathcal{B}\mathcal{C}}$ in (d) required for the computation of the TEE of the FCHI with OBC in the $x$ direction and PBC in the $y$ and $z$ directions. Here, the system has $2\times 3 \times 5$ unit cells and the subsystems are rotated compared to Fig.~\ref{fig:SketchKitaevPreskill3D}(b) along the $y$ axis such that they are translation invariant in the $x$ direction. The fluctuations of the mean as a function of the number $N_{MC}$ of MC steps per run are much smaller than the statistical error.\label{fig:ConvergenceMC}}
\end{figure}

Here, we briefly explain how we obtain the estimates for the errors of our MC measurements.  For each MC measurement, we launched $N_{\mathrm{seed}}$ instances of the algorithm, with each instance having a distinct seed of the random number generator. Typically, we chose $N_{\mathrm{seed}} = 100$. For each seed, the algorithm was performed until the Metropolis chain had a length of $N_{MC}$ MC steps, after which we evaluated the average of each run separately. Then, we computed the final value and error of the MC measurement as the mean and standard deviation, respectively, of the collection of $N_{\mathrm{seed}}$ averages per run. Therefore, the final value is the average after $N_{\mathrm{seed}} \times N_{MC}$ total MC steps.

In practice we have observed that the error estimated in this way is much larger than the fluctuations of the mean value of the measurements as a function of the number of MC steps after the initial convergence phase. For instance, in Fig.~\ref{fig:ConvergenceMC} we show the evolution of the mean value of the $\swapAmp$ observable of Eq.~\eqref{SWAPAmp} for the four EE measurements required for the computation of the TEE of the FCHI, namely $S^{(2)}_{\mathcal{B}}$, $S^{(2)}_{\mathcal{A}\mathcal{B}}$, $S^{(2)}_{\mathcal{A}\mathcal{C}}$, $S^{(2)}_{\mathcal{A}\mathcal{B}\mathcal{C}}$ as defined in Eq.~\eqref{TEE} of the main text. We focus here on the OBC case, \ie OBC in the $x$ direction and PBC in the $y$ and $z$ directions. Therefore, we think that the statistical fluctuations computed in this way might overestimate the actual error of the final MC measurement.

\subsection{Technical data~\label{sec:AppendixRunTimes}}

For the convenience of anyone wishing to reproduce our results, we have summarized some technical data including the acceptance rate, the number of MC steps and the run time of all computations discussed in the main text in Table~\ref{tab:MCData}. The computations were performed for the most part on machines with CPUs of type Intel(R) Xeon(R) E5-2680 v2 @ 2.80GHz (Ivybridge), with between 500 and 1000 cores in use simultaneously.

\begin{table}
	\begin{ruledtabular}
		\begin{tabular}{|l|l|l|l|l|l|l|l|l|}
			Physical observable & System size & Boundary conditions & MC observable & $r_{\text{update}}$ & \parbox{1.6cm}{Acceptance rate [\%]} & MC Steps & \parbox{1.6cm}{Total CPU hours}\\
			\hline		
			\multirow{4}{*}{$S^{(2)}_{\mathcal{A}_{N_x, N_y, N_{z, \mathcal{A}}}}$ in Fig.~\ref{fig:EEChiralHingeTi}(a)} & $2 \times 2 \times 20$ & OBC $\times$ OBC $\times$ PBC & $\myexp{\swapAmp}$ & 2 & $3.6$ & $10^8$ & 115500\\
			 & $2 \times 2 \times 20$ & OBC $\times$ OBC $\times$ PBC & $\myexp{\swapPhase}$ & 2 & $3.7$ & $10^7$ & 25000\\						
			 & $3 \times 2 \times 20$ & OBC $\times$ OBC $\times$ PBC & $\myexp{\swapAmp}$ & 2 & $2.1$ & $1.2 \times 10^8$ & 383370\\
			 & $3 \times 2 \times 20$ & OBC $\times$ OBC $\times$ PBC & $\myexp{\swapPhase}$ & 2 & $2.2$ & $10^7$ & 62854\\
			 \hline
			\multirow{2}{*}{$\var(\mathcal{A}_{N_x, N_y, N_{z, \mathcal{A}}})$ in Fig.~\ref{fig:EEChiralHingeTi}(b)} & $2 \times 2 \times 20$ & OBC $\times$ OBC $\times$ PBC & $\var(M_{\mathcal{A}})$ & 3 & $11$ & $8 \times 10^7$ & 43000\\									
			 & $3 \times 2 \times 20$ & OBC $\times$ OBC $\times$ PBC & $\var(M_{\mathcal{A}})$ & 3 & $11$ & $8 \times 10^7$ & 147300\\	
			\hline									
			\multirow{6}{4cm}{Overlap matrix $\mathcal{O}$ with eigenvalues shown in Fig.~\ref{fig:SketchKitaevPreskill3D}(a)} & $2 \times 2 \times 2$ & PBC $\times$ PBC $\times$ PBC & $\mathcal{O}_{\psi_1,\psi_2}^{1, abs}$, $\mathcal{O}_{\psi_1,\psi_2}^{2, abs}$ & $\infty$ & $18$ & $1 \times 10^7$ & 1385\\
			& $2 \times 2 \times 2$ & PBC $\times$ PBC $\times$ PBC &  $\mathcal{O}_{\psi_1,\psi_2}^{ phase}$ & $\infty$ & $18$ & $1 \times 10^7$ & 900\\
			 & $3 \times 3 \times 3$ & PBC $\times$ PBC $\times$ PBC & $\mathcal{O}_{\psi_1,\psi_2}^{1, abs}$, $\mathcal{O}_{\psi_1,\psi_2}^{2, abs}$ & $\infty$ & $8$ & $1 \times 10^7$ & 2500\\		
			 & $3 \times 3 \times 3$ & PBC $\times$ PBC $\times$ PBC & $\mathcal{O}_{\psi_1,\psi_2}^{ phase}$ & $\infty$ & $8$ & $1 \times 10^7$ & 2500\\
			 & $4 \times 4 \times 4$ & PBC $\times$ PBC $\times$ PBC & $\mathcal{O}_{\psi_1,\psi_2}^{1, abs}$, $\mathcal{O}_{\psi_1,\psi_2}^{2, abs}$ & $\infty$ & $7$ & $1 \times 10^7$ & 7000\\
			 & $4 \times 4 \times 4$ & PBC $\times$ PBC $\times$ PBC & $\mathcal{O}_{\psi_1,\psi_2}^{ phase}$ & $\infty$ & $7$ & $1 \times 10^7$ & 7000\\
			\hline	
			\multirow{4}{4cm}{Topological entanglement entropy $\gamma$ from Kitaev-Preskill cut in Fig.~\ref{fig:SketchKitaevPreskill3D}(b)} & $3 \times 3 \times 2$ & PBC $\times$ PBC $\times$ PBC & $\myexp{\swapAmp}$ & $\infty$ & $1$ & $1.6 \times 10^9$ & 112182\\
			 & $3 \times 3 \times 2$ & PBC $\times$ PBC $\times$ PBC & $\myexp{\swapPhase}$ & $\infty$ & $1$ & $10^8$ & 14312\\
			 & $3 \times 3 \times 3$ & PBC $\times$ PBC $\times$ PBC & $\myexp{\swapAmp}$ & $\infty$ & $1$ & $3.6 \times 10^9$ & 331495\\
			 & $3 \times 3 \times 3$ & PBC $\times$ PBC $\times$ PBC & $\myexp{\swapPhase}$ & $\infty$ & $1$ & $10^8$ & 14312\\	
			 \hline
			\multirow{4}{4cm}{Topological entanglement entropy $\gamma$ from rotated Kitaev-Preskill cut} & $2 \times 3 \times 5$ & PBC $\times$ PBC $\times$ PBC & $\myexp{\swapAmp}$ & $\infty$ & 1.7 & $1.4 \times 10^9$ & 142341 \\
			 & $2 \times 3 \times 5$ & PBC $\times$ PBC $\times$ PBC & $\myexp{\swapPhase}$ & $\infty$ &  2 &  $2 \times 10^8$ & 56728 \\
			 & $2 \times 3 \times 5$ & OBC $\times$ PBC $\times$ PBC & $\myexp{\swapAmp}$ & $\infty$ & 1 & $4.4 \times 10^9$ & 438886 \\
			 & $2 \times 3 \times 5$ & OBC $\times$ PBC $\times$ PBC & $\myexp{\swapPhase}$ & $\infty$ & 1.2 & $2 \times 10^8$ & 37083 \\
		\end{tabular}
	\end{ruledtabular}
	\caption{Overview of technical data of all MC runs whose results are discussed in the main text. Here, the MC observables are defined in Appendix~\ref{sec:AppendixMC}. As introduced in Appendix~\ref{sec:AppMCUpdate}, the parameter $r_{\text{update}}$ controls the maximal range of the spin exchange in the MC update, which can take a finite integer value or the value $\infty$ (meaning that there is no restriction on the maximal range). \label{tab:MCData}}
\end{table}

\end{document}